\documentclass[twocolumn]{aastex701}
\usepackage{amsmath}
 \usepackage{enumitem}
\usepackage{xcolor} 
\usepackage{amssymb}

\usepackage{mathtools}

\long\def\symbolfootnote[#1]#2{\begingroup%
\def\thefootnote{\fnsymbol{footnote}}\footnote[#1]{#2}\endgroup} 

\makeatletter\def\Hy@Warning#1{}\makeatother

\begin{document}

\title{Golden and Silver Dark Sirens for precise $H_0$ measurement with HETDEX}

\author[orcid=0000-0002-6689-8680,sname='North America']{Yixuan Dang}
\affiliation{Institute for Gravitation and the Cosmos, The Pennsylvania State University, University Park, PA 16802, USA}
\affiliation{Department of Physics, The Pennsylvania State University, University Park, PA 16802, USA}
\email[show]{ykd5167@psu.edu}  
\author[orcid=0000-0001-6932-8715,sname='North America']{Ish Gupta}
\affiliation{Institute for Gravitation and the Cosmos, The Pennsylvania State University, University Park, PA 16802, USA}
\affiliation{Department of Physics, University of California, Berkeley, CA 94720, USA}
\affiliation{Department of Physics and Astronomy, Northwestern University, 2145 Sheridan Road, Evanston, IL 60208, USA}
\affiliation{Center for Interdisciplinary Exploration and Research in Astrophysics (CIERA), Northwestern University, 1800 Sherman Ave, Evanston, IL 60201, USA}
\email[noshow]{ykd5167@psu.edu}  
\author[orcid=0000-0002-1328-0211,sname='North America']{Robin Ciardullo}
\affiliation{Department of Astronomy \& Astrophysics, The Pennsylvania State University, University Park, PA 16802, USA}
\affiliation{Institute for Gravitation and the Cosmos, The Pennsylvania State University, University Park, PA 16802, USA}
\email[noshow]{ykd5167@psu.edu}  
\author[0000-0002-2307-0146]{Erin Mentuch Cooper}
\affiliation{Department of Astronomy, The University of Texas at Austin, 2515 Speedway Boulevard, Austin, TX 78712, USA}
\email[noshow]{ykd5167@psu.edu}  
\author[orcid=0009-0003-5372-7318,sname='North America']{Shiksha Pandey}
\affiliation{Institute for Gravitation and the Cosmos, The Pennsylvania State University, University Park, PA 16802, USA}
\email[noshow]{ykd5167@psu.edu}  
\author[orcid=0000-0002-8925-9769,sname='North America']{Dustin Davis}
\affiliation{Department of Astronomy, The University of Texas at Austin, 2515 Speedway Boulevard, Austin, TX 78712, USA}
\email[noshow]{ykd5167@psu.edu}  
\author[orcid=0000-0002-2986-2371,sname='Asia']{Surhud More}
\affiliation{Inter-University Centre for Astronomy and Astrophysics, Pune, Maharashtra 411007, India}
\email[noshow]{ykd5167@psu.edu}

\author[orcid=0000-0002-5556-9873]{Rachel Gray}
\affiliation{SUPA, University of Glasgow, Glasgow, G12 8QQ, United Kingdom}
\email[noshow]{ykd5167@psu.edu}

\author[orcid=0000-0001-5403-3762]{Hsin-Yu Chen}
\affiliation{Department of Physics, The University of Texas at Austin, 2515 Speedway, Austin, TX 78712, USA}
\email[noshow]{ykd5167@psu.edu}

\author[0000-0003-2575-0652]{Daniel J. Farrow}
\affiliation{E. A. Milne Centre for Astrophysics
University of Hull, Cottingham Road, Hull, HU6 7RX, UK}
\affiliation{Centre of Excellence for Data Science,
Artificial Intelligence \& Modelling (DAIM),
University of Hull, Cottingham Road, Hull, HU6 7RX, UK}
\email[noshow]{ykd5167@psu.edu}

\author[orcid=0000-0001-6842-2371]{Caryl Gronwall}
\affiliation{Department of Astronomy \& Astrophysics, The Pennsylvania State University, University Park, PA 16802, USA}
\affiliation{Institute for Gravitation and the Cosmos, The Pennsylvania State University, University Park, PA 16802, USA}
\email[noshow]{ykd5167@psu.edu}  

\author[orcid=0000-0002-8434-979X,sname='North America']{Donghui Jeong}
\affiliation{Department of Astronomy \& Astrophysics, The Pennsylvania State University, University Park, PA 16802, USA}
\affiliation{Institute for Gravitation and the Cosmos, The Pennsylvania State University, University Park, PA 16802, USA}
\email[noshow]{ykd5167@psu.edu}  

\author[orcid=0000-0002-6186-5476]{Shun Saito}
\affiliation{Institute for Multi-messenger Astrophysics and Cosmology, Department of Physics, Missouri University of Science and Technology, 1315 N Pine Street, Rolla, MO 65409, USA}
\affiliation{Kavli Institute for the Physics and Mathematics of the Universe (Kavli IPMU, WPI), University of Tokyo, Chiba 277-8582, Japan}
\email[noshow]{ykd5167@psu.edu}

\author[orcid=0000-0001-7240-7449]{Donald P. Schneider}
\affiliation{Department of Astronomy \& Astrophysics, The Pennsylvania State University, University Park, PA 16802, USA}
\affiliation{Institute for Gravitation and the Cosmos, The Pennsylvania State University, University Park, PA 16802, USA}
\email[noshow]{ykd5167@psu.edu}

\author[orcid=0000-0003-3845-7586,sname='North America']{B. S. Sathyaprakash}
\affiliation{Institute for Gravitation and the Cosmos, The Pennsylvania State University, University Park, PA 16802, USA}
\affiliation{Department of Physics, The Pennsylvania State University, University Park, PA 16802, USA}
\affiliation{Department of Astronomy \& Astrophysics, The Pennsylvania State University, University Park, PA 16802, USA}
\email[noshow]{bss25@psu.edu}

\begin{abstract}
Gravitational waves (GWs) from compact binary coalescences are \emph{standard sirens} that provide a direct measure of the source's luminosity distance, enabling an independent measurement of the Hubble constant ($H_0$). While a \emph{bright siren}---a GW event with an identified electromagnetic (EM) counterpart---provided the first such constraint, most detections, currently dominated by black hole mergers, lack EM signatures. A measurement of $H_0$ is still possible with these \emph{dark sirens} by statistically associating GW events with galaxies in existing catalogs based on the sky localization. In this work, we explore the potential of two subsets of dark sirens categorized by their localization precision: 'golden' dark sirens, defined by a sky area localization $\leq 0.1 \text{ deg}^2$, and 'silver' dark sirens, which are more common but less precisely localized ($\leq 1 \text{ deg}^2$). Using the \emph{fifth internal data release} of the Hobby–Eberly Telescope Dark Energy Experiment (HETDEX), we assess the suitability of the Visible Integral-field Replicable Unit Spectrograph (VIRUS) for spectroscopic follow-up of dark sirens. VIRUS exposures of the standard HETDEX depth provide precise redshifts and exquisite completeness within $z =0.2$.
After a single year of observations with the upgraded LIGO-A$^\#$ network, the combined sample of golden and silver dark sirens with $z \leq 0.2$ at $H_0 = 70~\mathrm{km\,s^{-1}\,Mpc^{-1}}$ and follow-up VIRUS observations can potentially yield a few-percent constraint on $H_0$. Our predictions suggest that spectroscopic redshift surveys such as HETDEX can play a key role in realizing high-precision cosmology with dark sirens in the near future. 

\end{abstract}


\section{Introduction} 
It is remarkable, and deeply troubling that a century after Edwin Hubble’s discovery of the expansion of the universe, we still lack a precise and consistent determination of the rate of this expansion. The Hubble tension \citep[e.g.,][]{Verde:2019ivm,CosmoVerseNetwork:2025alb}---the $4$-$6\sigma$ discrepancy between the Planck Cosmic Microwave Background (CMB) value of the Hubble Constant \citep{Planck:2018vyg} and the local value of $H_0$ measured from the Supernova $H_0$ for the Equation of State of Dark Energy (SH0ES) program \citep[e.g.,][]{Riess:2021jrx,Riess:2025chq}---is one of the most pressing challenges in modern physics. The mismatch indicates either the presence of unknown systematics in the measurements or the existence of new physics \citep{Poulin:2018cxd,DiValentino:2019qzk,DiValentino:2021izs,CosmoVerseNetwork:2025alb}.

Gravitational waves (GWs) from compact binary coalescences (CBCs) provide a direct measurement of the source's luminosity distance, making them powerful \textit{standard sirens} for independent determinations of $H_0$ \citep{Schutz:1986gp,Holz:2005df}.  So-called \textit{bright sirens} constrain $H_0$ by allowing the direct detection of the event's electromagnetic (EM) counterpart and yield an unambiguous redshift for the source.  This approach was first demonstrated by the
binary neutron star (BNS) merger GW170817 \citep{collaboration_advanced_2015,abbott_gravitational-wave_2017} via the optical transient discovered near its host galaxy NGC~4993 \citep{abbott_multi-messenger_2017}. However, since GW170817, only one additional BNS event and a few neutron star–black hole (NSBH) mergers have been detected \citep{LIGOScientific:2025pvj}, none with a confirmed EM counterpart. The detectability of such counterparts depends on the intrinsic merger rate, the neutron-star equation of state (EoS), and the ability to localize and follow up events in a timely manner. While studies suggest that next-generation detectors could routinely identify BNS and NSBH mergers \citep{Bailes:2021tot,Branchesi:2023mws}, the number of events with observable counterparts is expected to remain small due to the small fraction of events with favorable source geometry (i.e., close to face-on or face-off systems), low merger rate \citep{KAGRA:2021duu,LIGOScientific:2025pvj}.

Fortunately, measurements of $H_0$ with GW sources are still possible despite the lack of EM counterparts. The statistical or dark-siren method \citep{Schutz:1986gp, gair_hitchhikers_2023, Gray:2021thesis, Gray:2019ksv} identifies potential host galaxies of CBCs by combining the GW sky localization with galaxy redshifts from existing spectroscopic or photometric catalogs \citep[e.g.,][]{Dalya:2018cnd,Dalya:2021ewn}. In this approach, all galaxies consistent with the GW localization are weighted by their probability of hosting the merger, and the resulting redshift distribution is combined with the GW-inferred luminosity distance to obtain a posterior on $H_0$.  One advantage of this approach is that rapid EM follow-up is not required; galaxy redshifts can be obtained well-after the time of the event.

Another approach—the spectral-siren method—uses population-level features in the \emph{source-frame mass distribution} of compact binaries \citep{Taylor:2012db, Mastrogiovanni:2021wsd, Ezquiaga:2022zkx}. Here “source-frame mass’’ refers to the true physical masses of the binary components, before cosmological redshift stretches the GW frequency and causes the \emph{detector-frame masses} to appear larger. Because GW detectors measure only redshifted masses, a mass–redshift degeneracy arises: a distant system with larger intrinsic masses can mimic a nearby system with smaller masses. Population-level modeling of the intrinsic mass distribution can partially break this degeneracy and thereby constrain $H_0$. Other methods that use GW observations to infer the Hubble constant include the stochastic-siren technique \citep{Cousins:2025bas}, the love siren method \citep{Messenger:2011gi,Chatterjee:2021xrm,Dhani:2022ulg}, and the cross-correlation of GW sources with the large-scale distribution of galaxies \citep{Namikawa:2015prh, Mukherjee:2019wcg,Mukherjee:2020hyn,Bera:2020jhx, Mukherjee:2022afz,Afroz:2024joi,Ghosh:2023ksl,Ghosh:2025qwc}.

In this paper, we focus on two subclasses of dark sirens: (1) \textit{golden dark sirens} \citep{Borhanian:2020vyr,Gupta:2022fwd,Chen:2025qsl,Zhan:2025jqg}, for which only a single cataloged galaxy is a plausible host, and (2) \textit{silver dark sirens}, for which multiple galaxies within a redshift catalog could potentially host the source. For the purpose of planning for EM observations, we adopt a practical observational classification: any GW event with a 90\% credible sky area smaller than $0.1~\mathrm{deg}^2$ is designated as a \textit{golden} dark siren, while those with a 90\% area below $1~\mathrm{deg}^2$ are classified as \textit{silver} dark sirens.  The advantage of these designations is that the classification is based on the GWs themselves, as the true number of potential host galaxies is unknown until dedicated follow-up observations are performed. 

Since both golden and silver dark sirens rely heavily on knowing the redshifts of the potential host galaxies near the position of the GW event, it is important to have a catalog that ensures a nearly complete detection of objects in the expected volume.  Moreover, previous works have established that for well-localized events (e.g., $\Delta\Omega_{90} \approx 1~\mathrm{deg}^2$), spectroscopic redshifts offer a 15\% better constraint on $H_0$ compared to photometric redshifts \citep{cross-parkin_dark_2025}.  Therefore, to fully exploit the potential of dark sirens, it is necessary to identify a facility capable of obtaining deep spectroscopy of all possible hosts of a GW event. The Visible Integral-field Replicable Unit Spectrograph \citep[VIRUS;][]{Hill2021} on the Hobby-Eberly Telescope \citep[HET;][]{Ramsey1998} is an ideal instrument for dark siren follow-up. The Hobby-Eberly Telescope Dark Energy Experiment \citep[HETDEX;][]{het_descrip} illustrates the power of VIRUS, as the instrument has recently been used to measure large scale structure over 540~deg$^2$ of the $1.9 < z < 3.5$ universe. In this work, we explore the potential of constructing HETDEX-style surveys for golden and silver dark sirens. Such a dedicated spectroscopic follow-up strategy has already been demonstrated for one well-localized dark siren event \citep{DarkEnergySurveyGravitationalWave:2025ykv}, motivating a systematic exploration for optimizing future observations. VIRUS is explored as a new avenue for dark siren follow-up, as its wide field of view and high multiplexing capability make it well-suited for efficiently obtaining spectroscopic redshifts of potential host galaxies within the sky localization region of a GW event.

The rest of the paper is structured as follows: In Section \ref{sec:event-specific} we introduce the idea of building event-specific catalogs with HET+VIRUS to accomplish the measurements required for GW sources detected with a network consisting of LIGO, Virgo and LIGO-India observatories.  This analysis validates the EM follow-up strategy.  Section \ref{sec:stats} presents the statistical framework used for the $H_0$ analysis performed, together with potential systematic effects and their mitigation. In Section \ref{sec:data}, we describe the specific models and data; Sections \ref{sec:results} and \ref{sec:conclusion} present results and conclusions.

\section{Toward Event-Specific Catalogs}
\label{sec:event-specific}

In this section, we outline the observational and detector requirements needed to construct \emph{event-specific} spectroscopic catalogs for well-localized GW sources. We first describe the capabilities of the Hobby–Eberly Telescope and the VIRUS integral-field spectrograph, which together provide the depth and completeness necessary for galaxy redshift follow-up over degree-scale localization regions. We then summarize the GW detector networks and sensitivity configurations assumed throughout this work, and finally validate the proposed EM follow-up strategy by quantifying the expected number of golden and silver dark sirens, the suitability of HETDEX-style observations, and the resulting constraints on $H_0$.

\subsection{VIRUS on the Hobby-Eberly Telescope}
\label{sec:het}
One of the most used instruments on the Hobby-Eberly Telescope is VIRUS, a low-resolution integral-field unit (IFU) spectrograph designed for wide-field surveys \citep{Hill2021}.  Since VIRUS is an IFU spectrograph, the instrument records the spectrum of every object falling onto its field of view without any pre-selection, and, since the instrument is coupled to a 10-m class telescope, redshifts well-beyond $z \sim 0.2$ can be accurately measured with exposure times of $\lesssim 20$~minutes.  Most importantly, VIRUS has the widest field-of-view of any IFU in existence, covering more than 55~arcmin$^2$ in a single observation.  It is therefore capable of surveying the entire 1~deg$^2$ sky area of a silver siren in $\sim 60$ pointings.

The fifth internal data release of HETDEX \citep{het_catalog_1} is particularly useful for exploring the utility of VIRUS for standard siren investigations. Although this database is primarily designed to identify emission lines brighter than $\sim 5 \times 10^{-17}$~ergs~cm$^{-2}$~s$^{-1}$ via the Emission Line eXplorer classification tool \citep[ELiXeR;][]{davis_hetdex_2023}, it also contains continuum-based redshifts for sources  brighter than $g \sim 23$ via the \texttt{Diagnose} spectral classification package \citep{Zeimann2024}. Roughly speaking, in just $\sim 21$~hours of exposure time, VIRUS can survey the entire 1 deg$^2$ localization area of a silver GW source at $z = 0.2$ and obtain a $\lesssim 0.15\%$ precision radial velocity for every potential host having a continuum luminosity more than half that of the Large Magellanic Cloud, or a star formation rate more than a tenth that of the Milky Way \citep{het_descrip, Hawkins2021, Zeimann2024}. In Section~\ref{sec:hetdex} we use the HETDEX catalog to conclude that any effect on $H_0$ due to incompleteness in the catalog is negligible.  This capability allows one to ignore most EM selections effects in our preliminary $H_0$ forecast.

\subsection{LIGO, Virgo and LIGO-India Detector Networks}
\label{sec:gw networks}
Next, we present a quantitative framework for designing HETDEX-style, event-specific spectroscopic surveys tailored to the golden and silver dark sirens expected from future ground-based GW detector networks. These networks will consist of upgraded versions of the currently operating LIGO and Virgo detectors, together with LIGO–India \citep{ligoindia}, which is now under construction. In our analysis, we consider three such network configurations, each operating under two progressively more sensitive instrument settings.  The first setting, the A$+$ configuration \citep{Aplus}, increases the accessible volume by a factor of $\sim 8$ relative to LIGO’s recently completed fourth observing run (O4) \citep{LIGOScientific:2025slb}; this is equivalent to an improvement in strain sensitivity by a factor of $\sim 2$.  The second, and the more ambitious A$^\#$ (pronounced A-sharp) configuration \citep{Asharp}, is expected to extend the detection volume by a factor of $\sim 60$ beyond O4. However the possibility of achieving A$^\#$ with Virgo remains uncertain. 

Hence we study the following network combinations in our analysis: 
\begin{itemize}[leftmargin=0pt]
    \item[] \textbf{HLV$+$}: which consists LIGO Hanford (H), LIGO Livingston (L), and Virgo (V) operating at A$+$ sensitivity; 
    \item[] \textbf{HLI$+$}: which uses LIGO Hanford, LIGO Livingston, and LIGO-India (I) at A$+$ sensitivity; and  
    \item[] \textbf{HLI\#}: which also utilizes LIGO Hanford, LIGO Livingston, and LIGO-India, but at A$^\#$ sensitivity. 
\end{itemize}

Our choice of detector networks is also guided by the current status of funded and scheduled upgrades: Virgo has a defined path toward Virgo+ sensitivity and LIGO–India is fully funded and under construction, offering a particularly advantageous geographical baseline for sky localization. Under the current sensitivity projections for KAGRA \citep{KAGRA:2020tym, KAGRA:2025dra}, its inclusion is not expected to significantly improve sky localization or distance inference for the well-localized events that dominate the golden and silver dark-siren sample. We note that any future improvement in KAGRA's sensitivity beyond current projections would only strengthen the conclusions of this work.

\subsection{Validation of the EM Follow-up Strategy}
\label{sec:em follow-up}
We analyze the validity of the proposed follow-up strategy, while considering three aspects of the problem. 
\begin{itemize}[leftmargin=0pt]
    \item []\textbf{(1) Expected detection rate of golden and silver dark sirens:}   We investigate the number of well-localized ($0.1~\mathrm{deg}^2$, golden) and moderately localized ($1~\mathrm{deg}^2$, silver) dark sirens that are expected within one year of observations under the A$+$ and A$^\#$ detector sensitivities. No golden dark sirens are predicted to be identified with the HLV$+$ or HLI$+$ networks, but 4 potential golden candidates per year are expected with the A$^\#$ sensitivity of HLI\#. In contrast, several silver dark sirens will be detectable per year with both HLV$+$ and HLI$+$, and these become commonplace with HLI\# (see Table.~\ref{tab:expected_number}). 
    \item[] \textbf{(2) Suitability of HETDEX-style surveys for follow-up dark siren signals:}   We assess whether the VIRUS instrument on the HET is suitable for building event-specific catalogs in terms of completeness and redshift precision. We show that HETDEX-style VIRUS observations will provide precise spectroscopic redshifts and detect all galaxies with $M_g\sim -17.6$ out to $z \approx 0.2$.  This sensitivity covers the redshift range of interest for near-future dark siren detections.   
    \item[]\textbf{(3) Expected precision on the Hubble constant ($H_0$) from one year of observations:}  We show that for sources at $d_L<980$ MPC, the error on $H_0$ inferred from the combination of $\sim 25$ golden and silver dark sirens under HLI\# can potentially be within a few percent.
\end{itemize}

Together, these results demonstrate the feasibility and robustness of the proposed strategy. The improved sensitivity curves for LIGO Hanford, LIGO Livingston, Virgo, and LIGO-India, in conjunction with the depth and redshift resolution achieved by HETDEX-style surveys, provide the key to precision cosmology with gravitational waves.

\section{Statistical Framework}
\label{sec:stats}
In this Section, we outline the statistical formalism used to infer the Hubble constant from a population of dark sirens with spectroscopic galaxy follow-up. We first introduce the Bayesian framework for constructing the posterior on $H_0$ given the GW measurements and the associated galaxy redshift catalog. We then describe how selection effects and host-galaxy weighting enter the likelihood.

\subsection{Bayesian Inference}
\label{sec:bayes}
Our analysis is conducted within a Bayesian statistical framework. Let $\{x\}$ denote the collection of $N_{\mathrm{GW}}$ gravitational-wave events, where $x^{(i)}$ (for $i=1, \dots, N_{\mathrm{GW}}$) represents the data for an individual event. Given the collection of GW data $\{x\}$ and the associated galaxy information $\{z, M\}$, the posterior distribution of $H_0$ is:

\begin{equation}
p\left(H_0 \mid \{x\}, \{z,M\}\right) \propto   \mathcal{L}\left(\{x\} \mid H_0, \{z,M\}\right) \times   \pi(H_0),
\label{eq:H0posterior}
\end{equation}

where $\mathcal{L}$ denotes the joint likelihood of the GW data for the set of dark sirens, and $\pi(H_0)$ is the prior on the Hubble constant, which we take to be uniform in the interval $[60, 80]~\text{km s}^{-1}\text{ Mpc}^{-1}$. While we adopt an $H_0$ prior range commensurate with the expected precision of future networks, we have verified that our primary conclusions are insensitive to this choice. A sensitivity test using a broad prior of $[40, 120] \text{ km s}^{-1} \text{ Mpc}^{-1}$ yielded consistent results for our high-statistics samples (e.g., silver sirens in HLI\#), indicating the posteriors are likelihood-dominated. For realizations with very few events ($N < 5$), the posterior’s width and shape is more sensitive to the prior; we therefore focus our analysis on the larger populations where the inference is robust.

In this work, we adopt two key assumptions that allow a clear statistical formulation of the dark–siren likelihood. First, we assume that the sample of galaxy redshifts is magnitude–complete within the GW localization volume, and reaches deep enough so that all galaxies that could plausibly host the event are included in the dataset. This is ensured for the mock data challenge that all true hosts have $M_g<-17.6$, which is the faintest absolute magnitude that can be detected at $z=0.2$. Second, we treat the spectroscopic redshifts as unbiased and effectively exact for our purpose. The validity of these assumptions for HETDEX-like VIRUS observations are further discussed in Sec.\,\ref{sec:hetdex}. 

For each event $i$, we consider a set of $N^{(i)}$ potential host galaxies with spectroscopic redshifts $\{z^{(i)}_j\}$ and absolute magnitudes $\{M^{(i)}_j\}$, where $j=1, \dots, N^{(i)}$ indexes the galaxies within the 90\% sky area. Under these assumptions, the joint likelihood factorizes naturally over events, and for each event the contribution to the likelihood is obtained by marginalizing over all galaxies that could host the source.

\begin{equation}
\begin{split}
    \mathcal{L}(\{ x \}| H_0,\{ z ,M\}) \propto \hspace{20pt} \\
    \prod ^{N_\mathrm{GW}}_i  \sum_j^{N^{(i)}}\frac{  \mathcal{L}(x^{(i)}\,|\,\Omega^{(i)}_{j}, d_{L}(z^{(i)}_j,H_0))}{\beta(H_0)} & p_\mathrm{Host}(j|\{ z^{(i)} ,M^{(i)}\},H_0).
\end{split}
\label{eq:bayes_extended_simplified}
\end{equation}

Here $\Omega^{(i)}_j,$ $j=1,\dots, N^{(i)},$ are the sky positions of the $N^{(i)}$ galaxies  within the 90\% credible region of the $i^{\it th}$ GW event, and the luminosity distance  $d_L(z^{(i)}_j, H_0)$ connects the EM-observed redshift to the GW-inferred distance through the Hubble constant $H_0$. The term $p_{\mathrm{Host}}$ encodes any desired luminosity weighting of host–galaxy probabilities, while the normalization term $\beta(H_0)$ captures GW selection effects, primarily detection efficiency, which are crucial for a self–consistent inference of $H_0$. 

We consider two choices for the weighting in our analysis: no weighting, where every galaxy has the same likelihood of being the GW host, and linear weighting, where the host probability is proportional to its absolute luminosity \citep{Gray:2021thesis}, i.e.,
\begin{equation}
\begin{aligned}
    p_\mathrm{Host}(j|\{ z^{(i)} ,M^{(i)}\},H_0)=\frac{L_j}{\Sigma_j^{N_\mathrm{i}} L_j}.
\end{aligned}
\end{equation}
For our main results we adopt the no-weighting scheme, since our mock universe is not generated using a luminosity-weighted selection and applying such weights would therefore be inconsistent with the  simulation model.


\subsection{Selection Effects}
\label{sec:selection effects}
A well-known selection effect in gravitational-wave astronomy is the analogue of the classical Malmquist bias, discussed in detail in \cite{Schutz:2011tw}. In GW astronomy, Malmquist-like bias encompasses all selection effects that make certain types of binaries easier to detect than others. Because GW detectors are amplitude-limited, systems that produce intrinsically stronger strain waveforms are overrepresented in the detected population. This includes biases due to the distance–inclination degeneracy as face-on (or face-off) binaries appear brighter than edge-on ones at the same distance; mass distribution as heavier binaries have larger horizon distances; spin-induced precession and higher harmonics as binaries with significant in-plane spins or large mass ratios generate richer waveform structure, enabling more precise parameter estimation; sky location relative to the detector network, which modulates the effective antenna response, and lastly astrophysical merger-rate evolution, which changes the relative number of detectable sources at different redshifts. All of these effects change the shape of the detected population, and all of them should be included in the full selection function $\beta(H_0)$. A rigorous numerical computation for the selection effects can be done in the most recent version of \texttt{gwcosmo} \citep{Mastrogiovanni:2023emh}.

The form of the GW selection effects for golden and silver dark sirens follows the framework set by \cite{chen_2_2018} and \cite{fishbach_standard_2019}, with the assumption that the catalog of galaxy redshifts is magnitude complete with no other biases:
\begin{equation}
\begin{split}
\beta(H_0)= \sum^{N\mathrm{gal}}_j P^\mathrm{GW}_{\mathrm{det}}(\Omega_j,d_L(z_j,H_0)) \\ \times \,p_\mathrm{Host}(j|\{ z^{(i)} ,M^{(i)}\},H_0)&
\end{split}
\end{equation}
where 
\begin{equation}
\begin{split}
    P^\mathrm{GW}_{\mathrm{det}}(\Omega_j,d_L(z_j,H_0)) = \\ \mathcal{L}(\mathrm{SNR}>12,  
     \Delta \Omega_{90} & <\Delta \Omega_{\rm threshold}|\Omega_j,d_L(z_j,H_0)).
\end{split}
\end{equation}
where SNR is the signal-to-noise of the detection, $N\mathrm{gal}$ is the number of galaxies in the catalog, and $\Delta \Omega_{\rm threshold} = 0.1$ deg$^2$ for golden dark sirens, and $1$ deg$^2$ for silver dark sirens. 

\begin{figure}
    \centering
    \includegraphics[width=\linewidth]{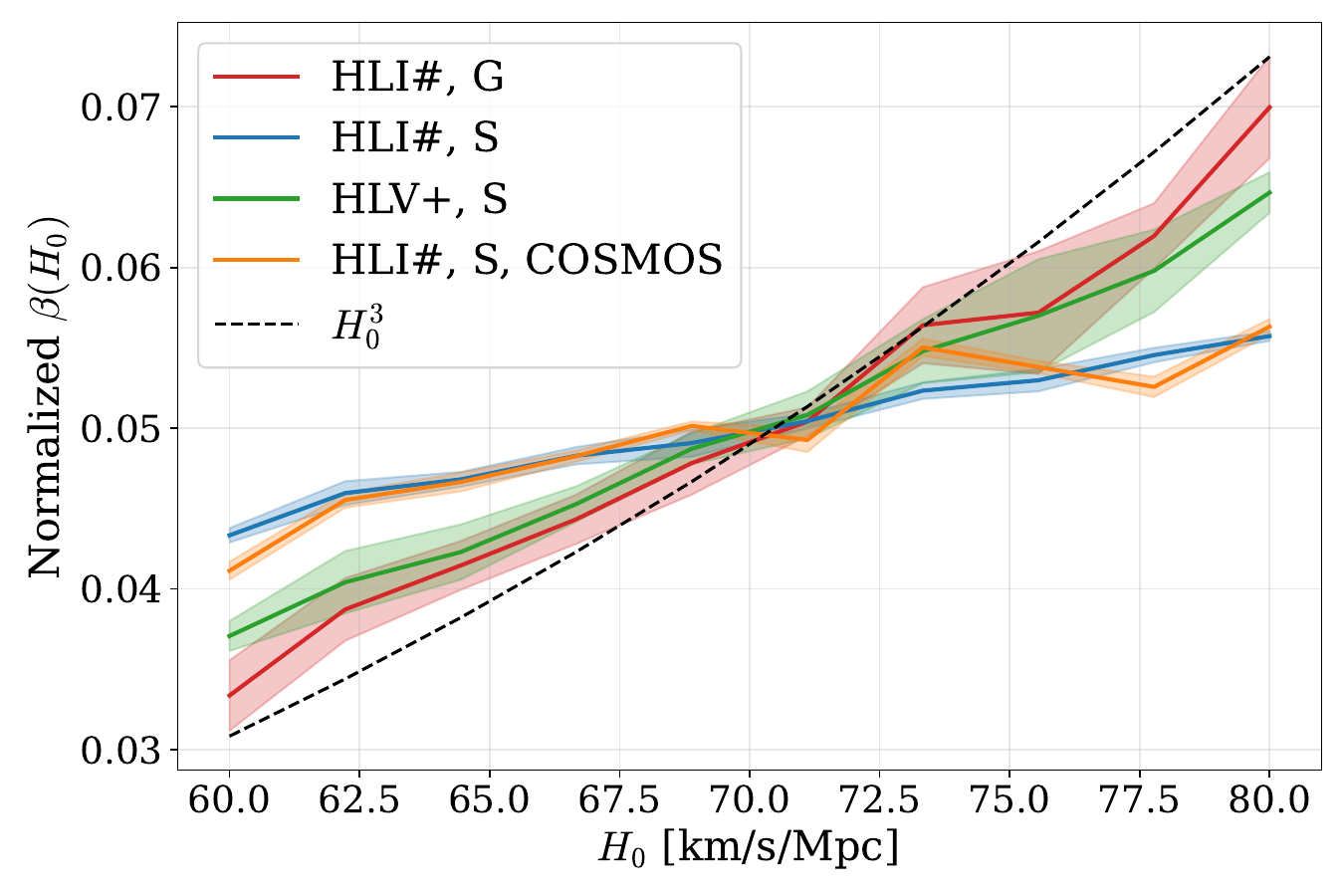}
    \caption{Numerically computed selection function shown in solid lines for HLI\# and HLV$+$, where G represents golden dark sirens with $\Delta \Omega_{90}\leq 0.1$\,deg$^2$ and S represents silver dark sirens with  $\Delta \Omega_{90}\leq 1$\,deg$^2$. For more accurate $H_0$ inferences later, we also computed a selection function for silver dark sirens within the HLI\# network, using the COSMOS field as the fiducial sky background. The shaded region shows the Monte Carlo uncertainty resulted from 5 realizations.}
    \label{fig:select func}
\end{figure}

Figure~\ref{fig:select func} shows the numerically computed selection functions for golden and silver dark sirens for the HLI\# and HLV$+$ detector networks. The selection function $\beta(H_0)$ is computed numerically using a Monte Carlo procedure based on simulated gravitational-wave injections generated with \texttt{GWBENCH} \citep{Borhanian:2020ypi}. \texttt{GWBENCH} is a fast Fisher-information–based forecasting software designed to benchmark the performance of GW detector networks. For each injection, we record the network SNR and  $\Delta\Omega_{90}$, and classify the event as detected if it satisfies $\mathrm{SNR} > 12$ and a localization threshold corresponding for golden ($\Delta\Omega_{90} \le 0.1~\mathrm{deg}^2$) to silver ($\Delta\Omega_{90} \le 1~\mathrm{deg}^2$) dark sirens. The maximum redshift is taken as $0.2$ at $H_0 = 70~\mathrm{km\,s^{-1}\,Mpc^{-1}}$ to remain consistent with the rest of the work. To obtain a smooth estimate of the detection probability as a function of sky location and luminosity distance, we construct three-dimensional kernel density estimators (KDEs) in $(\alpha, \delta, d_L)$ space—where $\alpha$ represents right ascension and $\delta$ represents declination—for both the full injection set and the detected subset, using identical kernels and bandwidths. The GW detection probability $P^{\mathrm{GW}}_{\mathrm{det}}(\Omega,d_L)$ is then evaluated as the ratio of the detected to total KDEs, which captures the relative detection efficiency as a function of geometry and distance. For simplicity, we draw Monte Carlo samples for $\Omega_j$ and $z_j$ uniformly over the sphere and comoving volume. This approach replaces the need to integrate over every galaxy within the mini-catalogs for each value of $H_0$. The procedure is repeated with different random seeds to estimate the Monte Carlo uncertainty. Finally, the resulting selection functions are normalized such that $\int \beta(H_0)\,\mathrm{d}H_0 = 1$, enabling direct comparison between detector networks and with the local-universe approximation, $\beta(H_0)\propto H_0^3$.

We find that the selection functions for HLI\# golden and HLV$+$ silver dark sirens closely follow the local-universe $\beta(H_0) \propto H_0^3$ scaling; this is consistent with the samples being dominated by nearby, high-SNR events. In contrast, the selection function for silver dark sirens detected with HLI\# exhibits a more pronounced deviation from the $H_0^3$ approximation. The difference in the selection function for silver dark sirens within the HLI\# network, relative to HLV$+$, is a direct consequence of the longer network baseline. The HLI\# configuration possesses significantly higher angular resolving power; consequently, its selection function for localized events deviates from one derived solely from a fixed SNR threshold. No golden dark sirens are found for the HLV$+$ configuration, as discussed in Section~\ref{sec:results} and hence the selection function for HLV$+$ golden dark sirens is not evaluated.

Finally, we derived a tailored selection function for silver dark sirens within the HLI\# network, using the COSMOS field as the fiducial sky background. This specific computation was necessitated by the small volume of the COSMOS field, where the standard assumption of large-scale homogeneity is expected to break down. We performed a parallel analysis for the significantly larger SHELA field, where there were no substantial deviations from the general isotropic selection function. Consequently, we adopted the generalized selection function for the SHELA analysis while maintaining the field-specific correction for COSMOS.

\section{Methods and Data}
\label{sec:data}

To test the robustness of the statistical framework, we first identify the golden and silver dark sirens from all events detectable in a year under each network, using the population model described below. A full Bayesian parameter estimation is performed for each of these golden and silver events. The distribution of galaxies is extracted from the fifth internal release of HETDEX. 

\subsection{GW Population Modeling and Inference}

Our population analysis is carried out with \texttt{GWBENCH} \citep{Borhanian:2020ypi}. \texttt{GWBENCH} provides rapid estimates of sky localization, distance uncertainties, and parameter-estimation accuracy for compact-binary signals without requiring full Bayesian inference, making it well suited for population-level studies and survey design. For the purposes of this work, it enables us to efficiently model the detection and localization capabilities of future GW networks and to propagate these effects into predictions for dark-siren cosmology.

For this analysis we assume that the mass distribution of binary black holes (BBHs) follows the \texttt{POWER LAW + PEAK} model examined in the third Gravitational-Wave Transient Catalog\symbolfootnote[2]{Although the most recent GWTC–4.0 catalog \citep{LIGOScientific:2025pvj} provides updated models for the mass distributions of black holes, this work was already well advanced when those results became available and a full reanalysis was not feasible. However, the revisions to the mass distribution are modest and do not materially affect our forecasts or alter any of our conclusions. For example, the local merger rate used in this work still lies within the newly predicted range.} GWTC-3.0 \citep{LIGOScientific:2018jsj, KAGRA:2021duu}, 
where the primary mass $m_1$ follows a power law with slope $\alpha = -3.4$ between 
$m_{\min} = 5\,M_{\odot}$ and $m_{\max} = 87\,M_{\odot}$, and smoothly tapers down below the minimum by a window width $\delta_m = 4.8\,M_{\odot}$. A Gaussian component centered at 
$\mu_{\mathrm{peak}} = 34\,M_{\odot}$ with width $\sigma_{\mathrm{peak}} = 3.6\,M_{\odot}$ and 
fractional weight $\lambda = 0.04$ accounts for the excess of events near the pair-instability gap. 
The secondary mass is drawn via 
\[
    p(q) \propto q^{\beta},
\]
with $\beta = 1.1$, where $q = m_2 / m_1 < 1$ denotes the mass ratio. 

The spin magnitudes of the black holes, which are restricted to align with the orbital angular momentum, follow a beta distribution \citep{Wysocki:2018mpo}:
\begin{equation}
        p(\chi_i \,|\, \alpha_{\chi_i}, \beta_{\chi_i}) =
        \frac{\chi_i^{\alpha_{\chi_i}-1} \, (\chi_{\max} - \chi_i)^{\beta_{\chi_i}-1}}
        {\mathrm{B}(\alpha_{\chi_i}, \beta_{\chi_i}) \, \chi_{\max}^{\beta_{\chi_i} + \alpha_{\chi_i} - 1}},
\end{equation} 
where $\chi_i$ is the spin magnitude between $[0, \chi_{\max}=1]$, and 
$\mathrm{B}(\alpha_{\chi_i}, \beta_{\chi_i})$ is the beta function. 
This restriction excludes precession effects while modeling spins aligned with the orbital angular momentum. 

The simulated population contains $10^6$ BBH mergers distributed across five redshift bins—$[0.005, 0.1]$, $[0.1, 0.5]$, $[0.5, 1]$, $[1, 2]$, and $[2, 3]$—with roughly $2 \times 10^5$ injections per bin. This stratified sampling approach ensures sufficient coverage of the parameter space across the entire detection range, particularly at higher redshifts where detection probability is lower. To reflect a physical population, these injections are subsequently reweighted onto a fine-grained grid of 50 logarithmic steps following the \citep{madau_dickinson_2014} star-formation-rate history,
\begin{equation}
    \psi(z) = (1+z)^{\gamma} 
    \left[ 1 + \left( \frac{1+z}{1+z_p} \right)^{\kappa} \right]^{-1},
    \label{eq:madau_dickinson}
\end{equation}
with $(z_p,\, \gamma,\, \kappa) = (1.9,\, 2.7,\, 5.6)$~\citep{Madau:1996hu}. This functional form is used here directly as a phenomenological parametrization of the redshift evolution of the BBH merger-rate density. In this approach, the effects of formation-to-merger time delays are incorporated implicitly through the assumed redshift dependence. We normalize the above equation to ensure that the simulated detections reflect realistic source populations with a local merger rate density consistent with $23.9^{+14.3}_{-8.6} \ \mathrm{Gpc}^{-3}\,\mathrm{yr}^{-1}$ \citep{LIGOScientific:2020kqk}.

For the subset of events identified by \texttt{GWBENCH} as golden or silver dark sirens, we then transition from Fisher-matrix forecasts to a full Bayesian parameter–estimation analysis.  Such an analysis is essential because GW events often exhibit non-Gaussian posteriors, higher-mode contributions, and distance–inclination degeneracies that cannot be reliably captured by Fisher approximations.  By using a full Bayesian follow-up, we ensure that the key sources driving our $H_0$ sensitivity are treated with the highest-fidelity inference available.

The parameter estimation for the simulated golden and silver dark sirens is performed using the \texttt{bilby} Bayesian inference software \citep{bilby_paper}. We employ the \emph{relative binning} technique—a fast, accurate method for evaluating GW likelihoods without requiring full waveform recomputation \citep{relbin_bilby,relbin_cornish,Zackay:2018qdy}—together with the \texttt{IMRPhenomXHM} waveform model \citep{Garcia-Quiros:2020qpx}. This waveform family is particularly well suited to our analysis because it includes higher-order multipoles and models their influence consistently across the inspiral, merger, and ringdown, which is essential for accurately constraining the inclination and luminosity distance of these events.

\subsection{The HETDEX HDR5 data}
\label{sec:hetdex}

\begin{figure}[tbh!]
    \centering
    \includegraphics[width=1.0\linewidth]{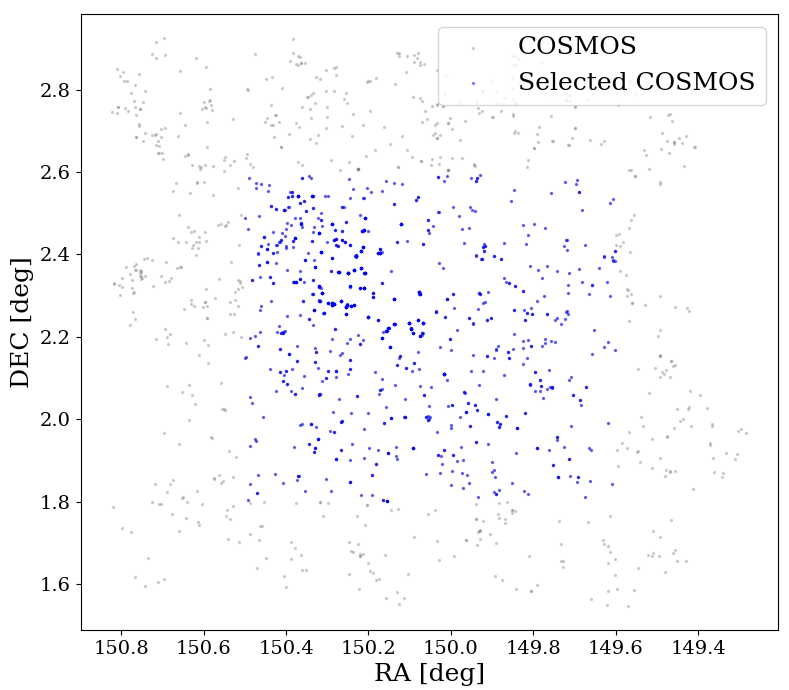}
    \caption{The distribution of galaxies in the COSMOS field. The blue points display those galaxies in the area selected for our mock data challenge with $z\leq0.2$ and apparent magnitude $g\leq22$.  In the region shown, the HETDEX fill-factor is close to unity.}
    \label{fig:cosmos}
\end{figure}

We draw our mock galaxies from the fifth internal data release of HETDEX\null. HETDEX is an untargeted spectroscopic survey designed to map the large-scale structure of Ly$\alpha$-emitting galaxies between $1.9 < z < 3.5$, and thereby place a constraint on the evolution of Dark Energy \citep{het_descrip}. To achieve this goal, the experiment uses the upgraded HET and VIRUS \citep{Hill2021}, a set of 78 IFU spectrographs, each delivering $R \sim 800$ spectra over the wavelength range $350\,\mathrm{nm} < \lambda < 550\,\mathrm{nm}$.  As each IFU covers $51\arcsec \times 51\arcsec$, VIRUS fills $\sim 22\%$ of the HET's $18\arcmin$ diameter field-of-view, and records a spectrum for every object whose light falls onto its fibers.  Among these objects are local galaxies with $z \leq 0.2$. 

The HETDEX survey consists of VIRUS observations in two large fields, a high Galactic latitude northern ``spring'' field and an equatorial ``fall'' field, and several smaller regions.  Because HETDEX was designed to measure the signal of baryonic acoustic oscillations at $z \sim 2.5$, its sky footprint is not ideal for our mock data challenge.  Specifically, the distribution of VIRUS IFUs on the HET focal plane is non-contiguous, and their orientation on the sky depends both on declination and whether a field was observed during its ``east'' or ``west'' track on the sky.  As a result, over most of the program's survey fields, the HETDEX footprint is irregular with a fill-factor, defined as the fraction of the survey’s geometric area that is actually covered by usable data, that is, slightly less than 22\%.  However, being a HETDEX science verification field, the COSMic Evolution Deep Survey (COSMOS) field \citep{Scoville:2006vq, Casey:2022amu} was observed by the survey multiple times, and as a result, it has a near-uniform fill-factor.  This is demonstrated by Figure~\ref{fig:cosmos}, which displays the positions of all of the field's HETDEX `galaxy' detections with $g \le 22$ and $z \le 0.2$.  The uniform spatial coverage, especially near the field center, is ideal for our dark siren tests. To avoid the irregular geometric shape on the edge and maximize the fill-factor, we have selected the central blue region labeled ``Selected COSMOS'' for the mock sky background. 

\begin{figure*}
    \includegraphics[width=\linewidth]{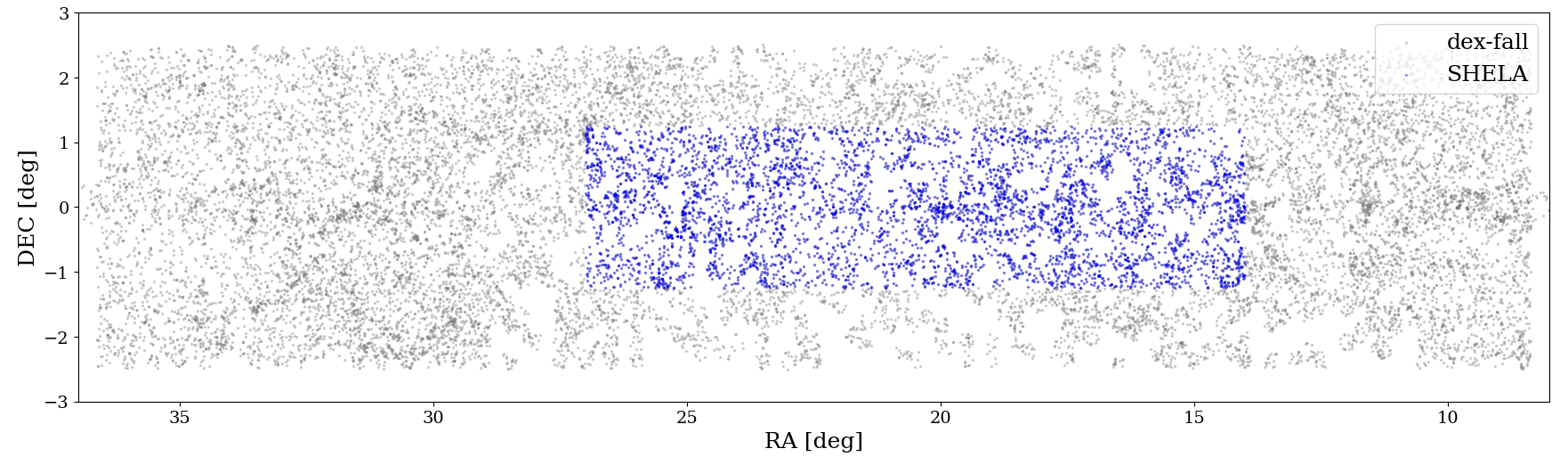}
    \caption{The distribution of galaxies in the HETDEX fall survey area. The blue points represent galaxies in the SHELA field with $z\leq0.2$ and apparent magnitude $g\leq22$.  The large void-like regions in the map are due to unobserved fields.  SHELA  
    provides a complementary data set to COSMOS in our mock data challenge:  its $\gtrsim 30$ larger area minimizes the effect of cosmic variance on our analysis, but its much lower fill-factor reduces the number of host galaxy candidates by a factor of $\sim 4.6$.}
    \label{fig:shela}
\end{figure*}

There is one drawback to using COSMOS in our dark siren challenge: because the field is relatively small, the effects of local large-scale structure will introduce more statistical fluctuations into the results (see Section~\ref{sec:results}).  Therefore, to assess the robustness of our calculations, we also use the Spitzer-HETDEX Exploratory Large-Area (SHELA) field \citep[a 24~deg$^2$ subset of the fall field;][]{SHELA_2016} in our dark siren analysis.  This region is shown in Figure~\ref{fig:shela}.  While SHELA has a far lower ($\sim 20\%$) fill-factor and a much more fragmented tiling pattern, its $\sim 30$ times larger area presents a useful complementary dataset.

\subsubsection{GW Mock Data Analysis}
For this study, we simulate GW events across the entire sky and then identify those that qualify as potential golden or silver dark sirens (i.e., those with $\Delta \Omega_{90}<0.1~\mathrm{deg}^2$ and $\Delta \Omega_{90}<1~\mathrm{deg}^2$) via a full Bayesian parameter estimation using \texttt{bilby} \citep{bilby_paper}. To emulate a realistic galaxy environment for each simulated event, we shift the sky coordinates of each simulated GW source so that it aligns with that of an $M_g < -17.6$ galaxy in the COSMOS or SHELA field, whose luminosity distance (computed assuming $H_0 = 70~\mathrm{km\,s^{-1}\,Mpc^{-1}}$) is closest—-typically within $0.5\%$—to that of the simulated merger. This procedure ensures that the spatial distribution of galaxies in the mock catalog is consistent with the true sky location of the simulated GW event and that the galaxy sample used in the $H_0$ inference is complete.

We select galaxies as potential hosts based on the credible sky volume of each GW injection, before passing the information in the catalog onto Eq.~(\ref{eq:bayes_extended_simplified}). A galaxy is selected as a potential host if it satisfies both of the following conditions: (1) its sky location falls within the 90\% credible region of the GW event and (2) its redshift is within the maximum redshift range corresponding to the 90\% luminosity distance posterior, assuming a Hubble Constant in the range $[\,60,80\,]\,\mathrm{km\,s^{-1}\,Mpc^{-1}}$. We have verified that extending the sky area to encompass the 99\% credible region leads to no noticeable change in the $H_0$ posterior. For a very small fraction of injections, potential host galaxies beyond $z=0.2$ are excluded because the mock universe extends to only $z=0.2$ despite the distance posterior falling within the allowed maximum redshift range. We have verified that this truncation has no discernible impact on the resulting $H_0$ posteriors. In future analyses, we will ensure that the prior range extends sufficiently beyond the intended redshift cutoff, so that the posterior support is not artificially limited by the bounds of the prior.

\subsubsection{Detection and Precision}
HETDEX detects objects using two different analysis pipelines.  The first, \texttt{ELiXer} \citep{davis_hetdex_2023}, identifies and classifies all emission lines with fluxes greater than $\sim 5 \times 10^{-17}\,\mathrm{erg\,cm^{-2}\,s^{-1}}$ (with the limit brightening to $\sim 40 \times 10^{-17}\,\mathrm{erg\,cm^{-2}\,s^{-1}}$ in the extreme blue limit of the experiment’s spectral range). The second, \texttt{Diagnose}, operates on objects detected in the $g$-band continuum, which encompasses most of VIRUS' spectroscopic bandpass.  This pipeline classifies objects based on their absorption lines and 
assigns redshifts to sources as faint as $g \sim 23$ \citep{Zeimann2024, het_catalog_1}. 

The HETDEX pipelines achieve redshift precisions ranging from $\sim 30\,\mathrm{km\,s^{-1}}$ for bright, high signal-to-noise ratio sources to $\lesssim 100\,\mathrm{km\,s^{-1}}$ for the faintest detections \citep{het_descrip,Hawkins2021}. At the luminosity distances relevant for this work ($300-1000$ Mpc), these uncertainties correspond to fractional redshift errors of $0.05\%$--$0.15\%$ in the best cases and $0.16\%$--$0.50\%$ for the worst cases (assuming $H_0=70~\mathrm{km\,s^{-1}\,Mpc^{-1}}$). We therefore consider the effect of redshift uncertainty negligible for the purposes of constraining $H_0$.

\subsubsection{Depth and Completeness}

HETDEX is designed primarily to detect and measure emission lines. Consequently, direct comparisons to surveys which target objects based on their continuum brightness are difficult. For example, \citet{het_descrip} have used artificial emission line experiments to characterize HETDEX completeness as a function of both emission-line flux and wavelength. However, there are few measurements in the literature of emission-line luminosity functions for comparison: the ones that exist either have much brighter flux limits, are severely compromised by sample size, pre-select targets on the basis of continuum brightness, or target the H$\alpha$ line, which is outside the HETDEX spectral range (see, e.g., \citealt{Salzer2000, Gallego2002, Ciardullo2013}). However, given that the continuum analysis program \texttt{Diagnose} reports a confident classification for $98.5\%$ of $g < 22$ detections, and $86.4\%$ of detections with $g < 23$ \citep{het_catalog_1}, it would seem likely that for $g < 22$ objects, the HETDEX catalog is close to complete.

\begin{figure}[!bth]
        \centering
        \includegraphics[width=0.95\linewidth]{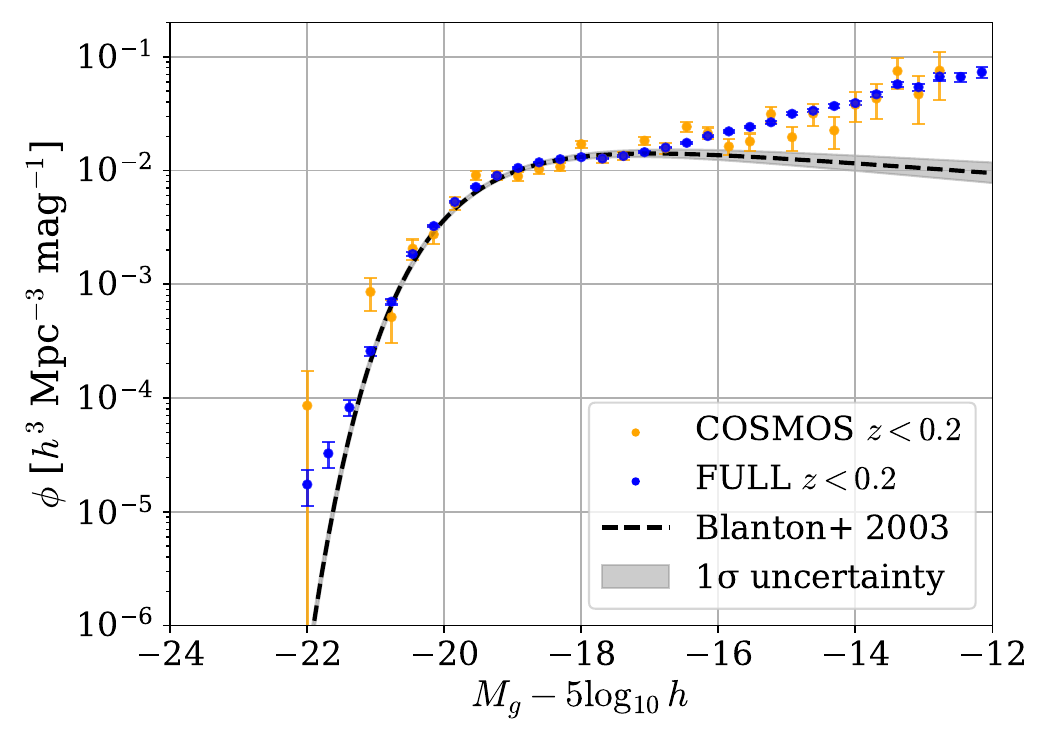}
        \label{fig:completeness}
        \caption{The $g$-band galaxy luminosity function derived from the HETDEX catalog compared to that found from the SDSS \citep{SDSS:2002vxn}. The orange points are derived solely from galaxies in the COSMOS field; the blue points represent galaxies from the full $79~\mathrm{deg}^2$ of the HETDEX DR5 catalog. The apparent magnitude limit is taken conservatively as $g\sim22$. The steep increase in the number of galaxies fainter than $M_g \sim -17$ is also seen in the DESI DR2 data \citep{moore2025desidr2galaxyluminosity}. } 
        \label{fig:LF}
\end{figure}
To confirm this assessment, we created a $g$-band luminosity function based on photometric magnitudes using the classic $V/V_{\rm max}$ method \citep{Schmidt:1968kn, Huchra1973}.  For this calculation, we do not use the pseudo-$g$ magnitudes derived from the HETDEX spectra; these data are often compromised by the combination of the limited field-of-view of each IFU and the non-contiguous distribution of IFUs across the HET's focal plane.  (This is especially true for extended objects which fall near an IFU edge.)  Instead, we use the Source Extractor-based isophotal magnitudes \citep{barbary2016} computed by ELiXeR \citep{davis_hetdex_2023} from the ancillary imaging surveys that HETDEX uses to assist in object classification; these data include $g$-band images from the Hyper Suprime-Cam Subaru Strategic Program \citep[HSC-SSP;][]{HSC-SSP}, the SHELA survey \citep{SHELAcat_2023}, and the Dark Energy Camera Legacy Survey \citep[DECaLS;][]{DECaLS}, among others. (A full list of the images, along with their depths is given in Table~1 of \citealp{davis_hetdex_2023}).  These data go much deeper in the continuum than HETDEX spectroscopy, and do not require extrapolations for parts of galaxies that fall outside the $51$\arcsec$\times51$\arcsec\ field of each VIRUS IFU.  In all cases, when multiple images are available, we use the deepest data, and only where no ancillary images are available do we revert to the pseudo-$g$-band magnitudes computed from HETDEX spectra (as described in \citealt{het_descrip} and \citealt{het_catalog_1}).

Figure~\ref{fig:LF} compares the luminosity function derived from the HETDEX data to the fitted Schechter function \citep{Schechter:1976iz} of SDSS \citep[$\phi_*=2.18\pm0.33$, $M_*-5\log_{10}h=-19.39\pm0.02$ and $\alpha=0.89\pm0.03$;][]{SDSS:2002vxn}. For the figure, all the $g$-band magnitudes described above have been corrected for foreground extinction using the line-of-sight reddening estimates from the \citet{schlafly2011} recalibration of the \citet{Schlegel1998} dust maps, using a standard $R_V = 3.1$ extinction law.  (In all cases, these corrections are quite small, as the HETDEX fields were originally selected to have minimal Galactic reddening.)  To be consistent with \cite{SDSS:2002vxn}, our absolute $g$-band magnitudes have been calculated using an averaged k-correction for each redshift bin obtained from the Galaxy And Mass Assembly data \citep[GAMA;][]{Driver:2022vyh}, and a luminosity density evolution of $Q*(z-0.1)$, where $Q$ is taken to be $2.04$. Overall, the agreement between the two measurements is excellent for absolute magnitudes brighter than $M_g \sim -18$.  At fainter magnitudes, the SDSS function is best fit by a flat power-law  with a faint slope of $\alpha \sim -0.9$.  Figure~\ref{fig:LF} displays this function, and its extrapolation beyond $M_g \sim -16$.  In contrast, the HETDEX luminosity function increases steeply at magnitudes fainter than $M_g \sim -18$, with a non-decreasing slope. This behavior is in excellent agreement with that derived by \cite{moore2025desidr2galaxyluminosity} for DESI-DR2 galaxies. 

The luminosity function of Figure~\ref{fig:LF} and its agreement with that found by DESI-DR2 presents a strong argument for the completeness of HETDEX\null.  Specifically, at $z\leq0.1$, the HETDEX catalog represents a volume-limited dataset for all galaxies more luminous than $M_g \sim -15.4$, i.e., a full magnitude fainter than the Small Magellanic Cloud.  Even at $z=0.2$, this limit only brightens to $M_g \sim -17.6$, a value similar to the absolute magnitude of the Large Magellanic Cloud.

Finally, we can compare our results to those found by \cite{Alfradique:2025tbj} using the \texttt{Magneticum} suite of cosmological simulations.  Using the simulation's mock galaxy dataset, the authors quantified parent galaxy detection completeness as a function of $r$-band apparent magnitude and found that, within $1600$~Mpc ($z = 0.307$ for $H_0 =70\,\mathrm{km\,s^{-1}\,Mpc^{-1}}$), an apparent magnitude cut of $r\sim22$ yields $98\%$ completeness.  This result is similar to what should be attained with VIRUS: although local galaxies can be up to a magnitude fainter in $g$ than in $r$ \citep[e.g.,][]{Bell2003}, our smaller redshift limit of $z \leq 0.2$ ($D_L \leq 980$~Mpc at $H_0 = 70~\mathrm{km\,s^{-1}\,Mpc^{-1}}$ more than compensates for this systematic effect.  Moreover, any residual incompleteness is further mitigated by weighting by galaxy properties: e.g., if host likelihood scales with star-formation rate, missing a few faint dwarf elliptical/spheroidal galaxies matters very little, while the emission-lines associated with star formation will increase VIRUS survey depth.  In short, HETDEX style-observations with VIRUS achieve over $98\%$ completeness in our distance range, so galaxy incompleteness should not significantly increase the $H_0$ error. We conclude that almost all potential galaxy hosts within $z=0.2$ will be captured by a dedicated VIRUS survey by HET.

\section{Results}
\label{sec:results}

\begin{figure}
    \centering
    \includegraphics[width=\linewidth]{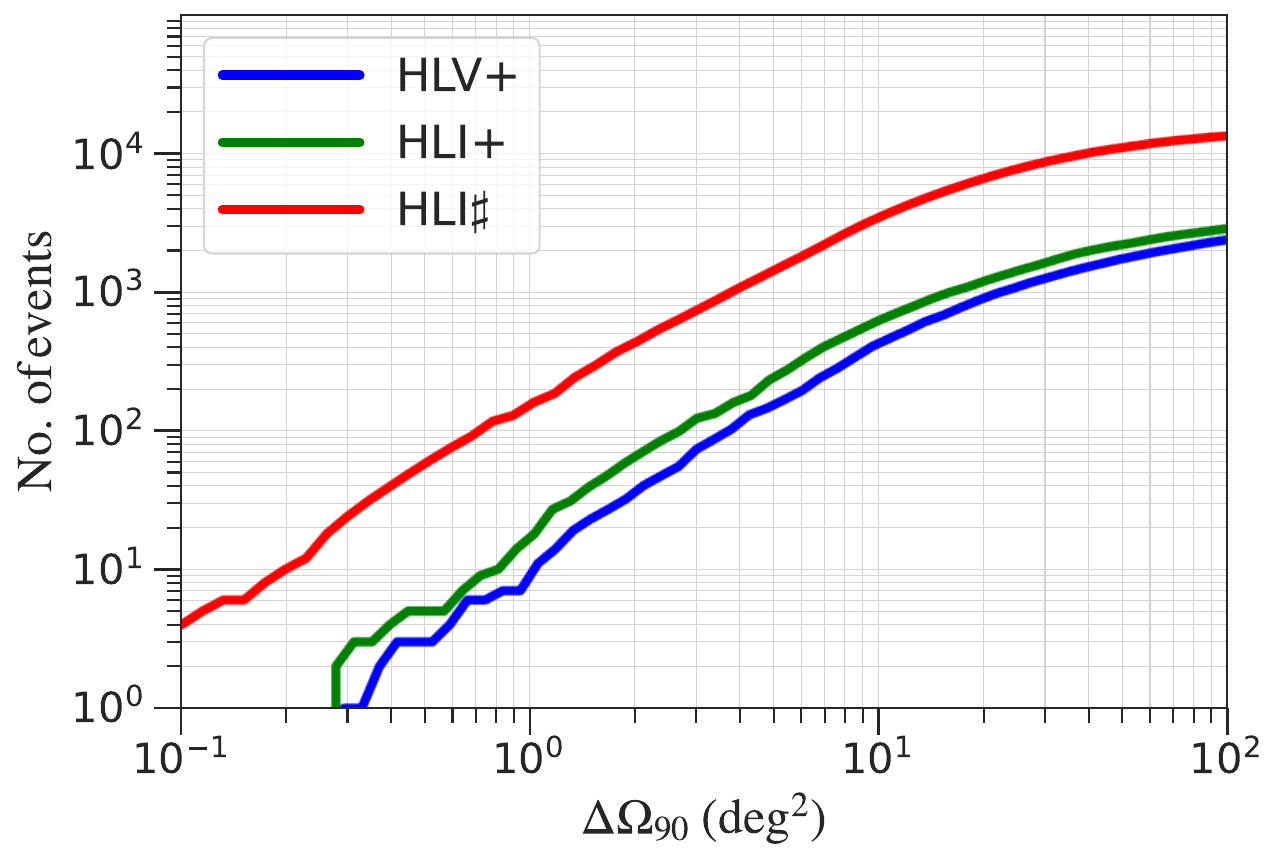}
    \caption{The expected cumulative number of GW events detected per year by the HLV$+$, HLI$+$, and HLI$\#$ networks as a function of their 90\% credible sky area.}
    \label{fig:gwbench}
\end{figure}
Figure~\ref{fig:gwbench} shows the \texttt{GWBENCH} predictions for the number of GW events expected per year as a function of their 90\% credible sky area for the three detector network configurations under consideration. Table~\ref{tab:expected_number} summarizes this information by giving the expected number of dark siren events classified as either golden or silver, with the addition of a 10-year forecast. The results show a clear improvement in sky localization with each upgrade in the detector sensitivity. We also notice that the HLI$+$ network detects more well-localized events than HLV$+$ due to the longer baseline between LIGO-India and LIGO Hanford or Livingston. Events with a 90\% credible sky area smaller than $1~\mathrm{deg}^2$ are detected in all of the networks considered, and the number of these silver events is much larger in the HLI\# network. More importantly, events with localizations below $\Delta\Omega_{90} < 0.1$~deg$^2$, i.e., the golden dark sirens, begin to frequent in HLI\#.  These numbers provide a robust estimate on the approximate number of follow-up events that would be available for an HET-like program, provided that they occur in the northern hemisphere, which happens at a near 50\% chance.

\begin{table}[!b]
\centering
\begin{tabular}{c|cc|cc}
\hline
\hline
& \multicolumn{2}{c|}{Per Year} & \multicolumn{2}{c}{10-Year Forecast} \\
\cline{2-5}
Network & $< 0.1 \deg^2$ & $< 1 \deg^2$ & $< 0.1 \deg^2$ & $< 1 \deg^2$ \\
\hline
HLV+    & 0 & 7   & 5 & 127 \\
HLI+    & 0 & 17  & 8 & 195 \\
HLI\#   & 4 & 136 & 48 & 1311 \\
\hline
\hline
\end{tabular}
\caption{Expected number of GW events per year and over a 10-year period with sky area localization thresholds corresponding to golden and silver dark sirens.}
\label{tab:expected_number}
\end{table}

\begin{figure*}
    \centering
    \includegraphics[width=0.9\linewidth]{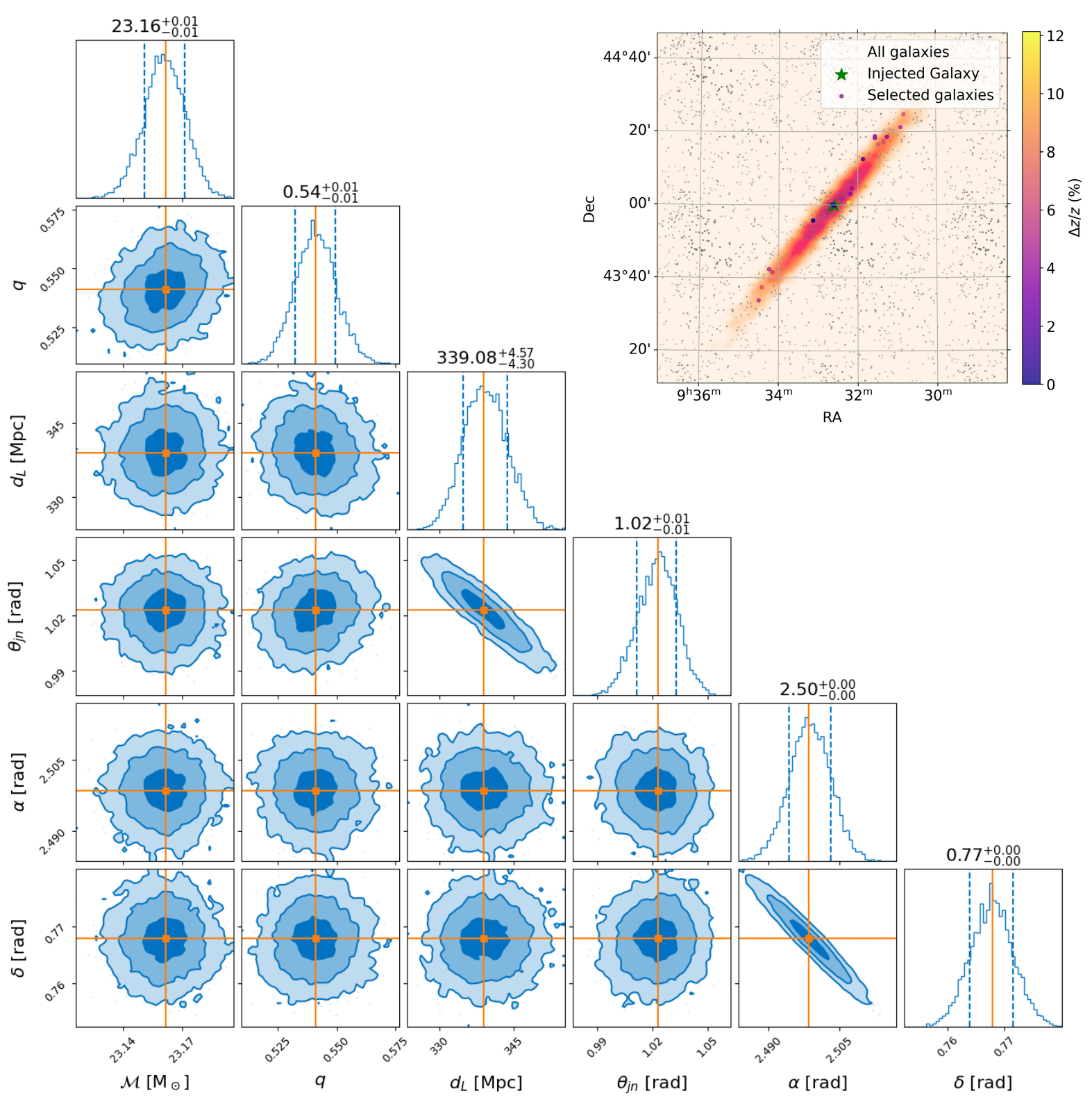}
    \caption{Example of a silver dark siren with $\Delta \Omega_{90}=0.18~\mathrm{deg}^2$. The corner plot shows the \texttt{bilby} posteriors results of chirp mass ($\mathcal{M}$), mass ratio ($q$), luminosity distance ($d_L$), inclination ($\theta_{\mathrm{JN}}$ as no precession is injected), Right Ascension (RA) and Declination (DEC). The sky map displays the 90\% credible region, along with the galaxies of the COSMOS field. The green star represents the injected galaxy. Galaxies that are identified as potential hosts are displayed as circles; these are the systems that contribute to the $H_0$ posterior. The color of each galaxy represents the fractional difference between its redshift and the redshift of the true galaxy. The fractional errors for the sky position are $+0.29536 / -0.28469$ (\%) in RA and $+0.80401 / -0.84067$ (\%) in Dec.}
    \label{fig:silver_inj}
\end{figure*}
\begin{figure*}
    \centering
    \includegraphics[width=0.9\linewidth]{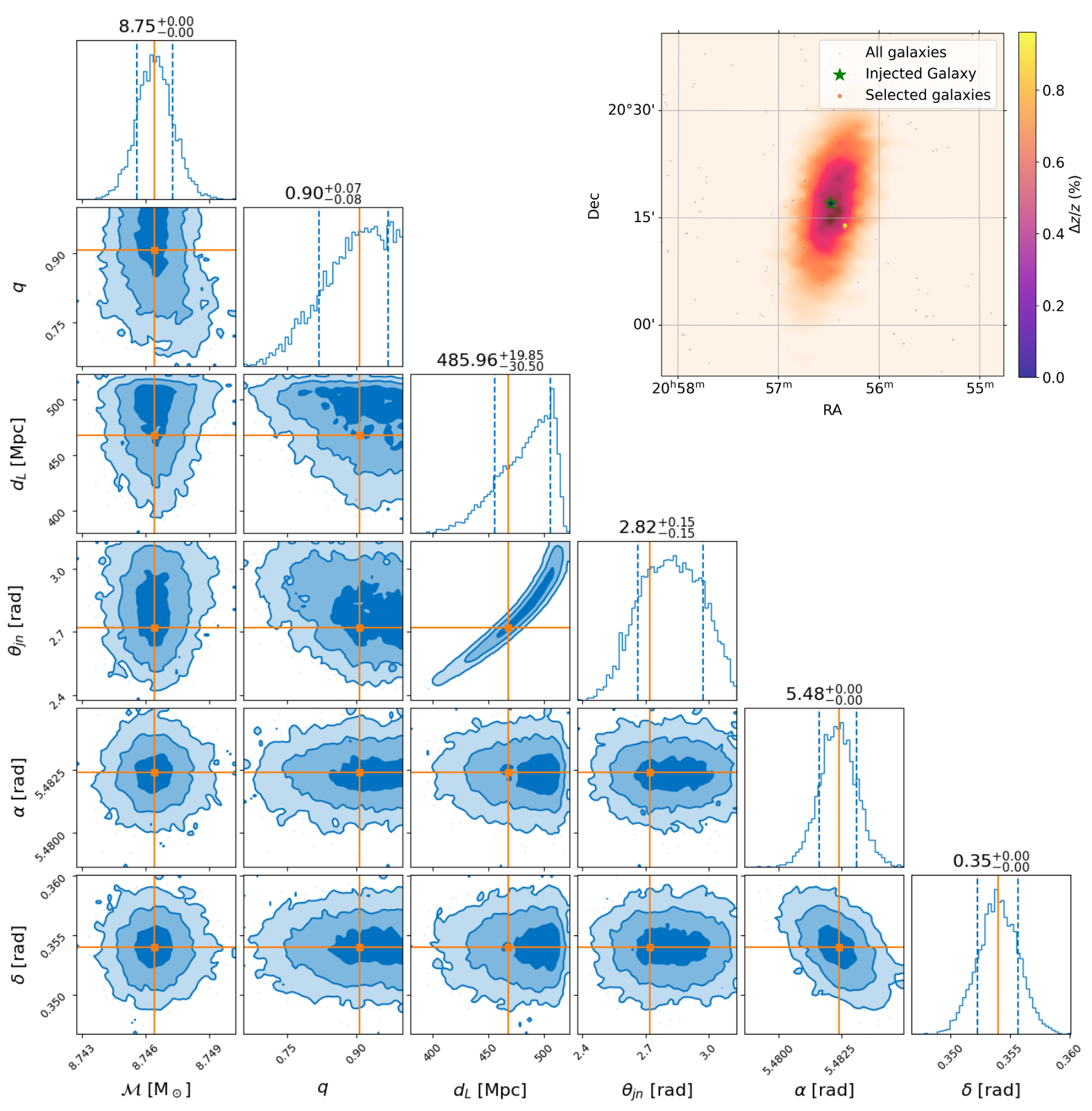}
    \caption{Example of a golden dark siren with $\Delta \Omega_{90}=0.07~\mathrm{deg}^2$. The corner plot shows the \texttt{bilby} parameter estimation results for the same set of parameters described in Fig.\,\ref{fig:silver_inj}; the sky map displays the 90\% credible region, along with galaxies in the SHELA field in HETDEX that have been filtered out as potential hosts. The green star represents the injected galaxy. The color of each galaxy represents the fractional difference between its redshift and the redshift of the true galaxy.} Only one galaxy (in purple, coincident with the injected galaxy) dominates the $H_0$ posterior;the other candidates (in yellow) deviate substantially in redshift.The fractional errors for the sky position are $+0.02158 / -0.02390$ (\%) in RA and $+0.78875 / -0.82616$ (\%) in Dec.
    \label{fig:golden_inj}
\end{figure*}

Our mock data analysis uses HETDEX data in the central 1~deg$^2$ region of the COSMOS field.  In area, this matches the upper limit for the definition of our silver dark sirens.  However, since GW sky localizations are frequently elongated with irregular geometries, a single event's 90\% region may extend beyond the limits of our data. To guarantee full coverage using our galaxy catalog, we tile the field by compactly replicating it along its edges, ensuring that the effective background area always exceeds the sky localization. The resulting  pattern is shown in the sky map in Fig.~\ref{fig:silver_inj}. In this figure, we also provide the full parameter estimation results for a representative silver dark siren. The corner plot illustrates the posterior distributions for the intrinsic and extrinsic parameters, including chirp mass ($\mathcal{M}$), mass ratio ($q$), and the sky coordinates. Notably, the degeneracy between the luminosity distance ($d_L$) and the binary inclination ($\theta_{\mathrm{JN}}$) is well-constrained, which is critical for reducing the distance uncertainty. The figure also illustrates the positional uncertainty of a typical silver dark siren with multiple possible host galaxies. This particular silver event uses the region of the COSMOS field where HETDEX has nearly $100\%$ spectroscopic coverage. Consequently, the example represents a realistic expectation with future detector networks.

Figure~\ref{fig:golden_inj} shows the corner plot and corresponding sky map for a representative golden dark siren in the SHELA field. In this example, a single galaxy (shown in purple) dominates the $H_0$ posterior. This event has relatively few potential hosts because the HETDEX HDR5 fill-factor of the region is relatively low (only $\sim 20\%$). Whether our observational criterion for identifying a ``golden'' dark siren (i.e., $\Delta\Omega_{90}\leq 0.1~\mathrm{deg}^2$) coincides with the definition commonly used in the literature (i.e., the presence of a single plausible host) depends sensitively on the galaxy density of the sky region in which the GW event occurs.


\begin{figure}[bt!]
    \centering
    \includegraphics[width=\linewidth]{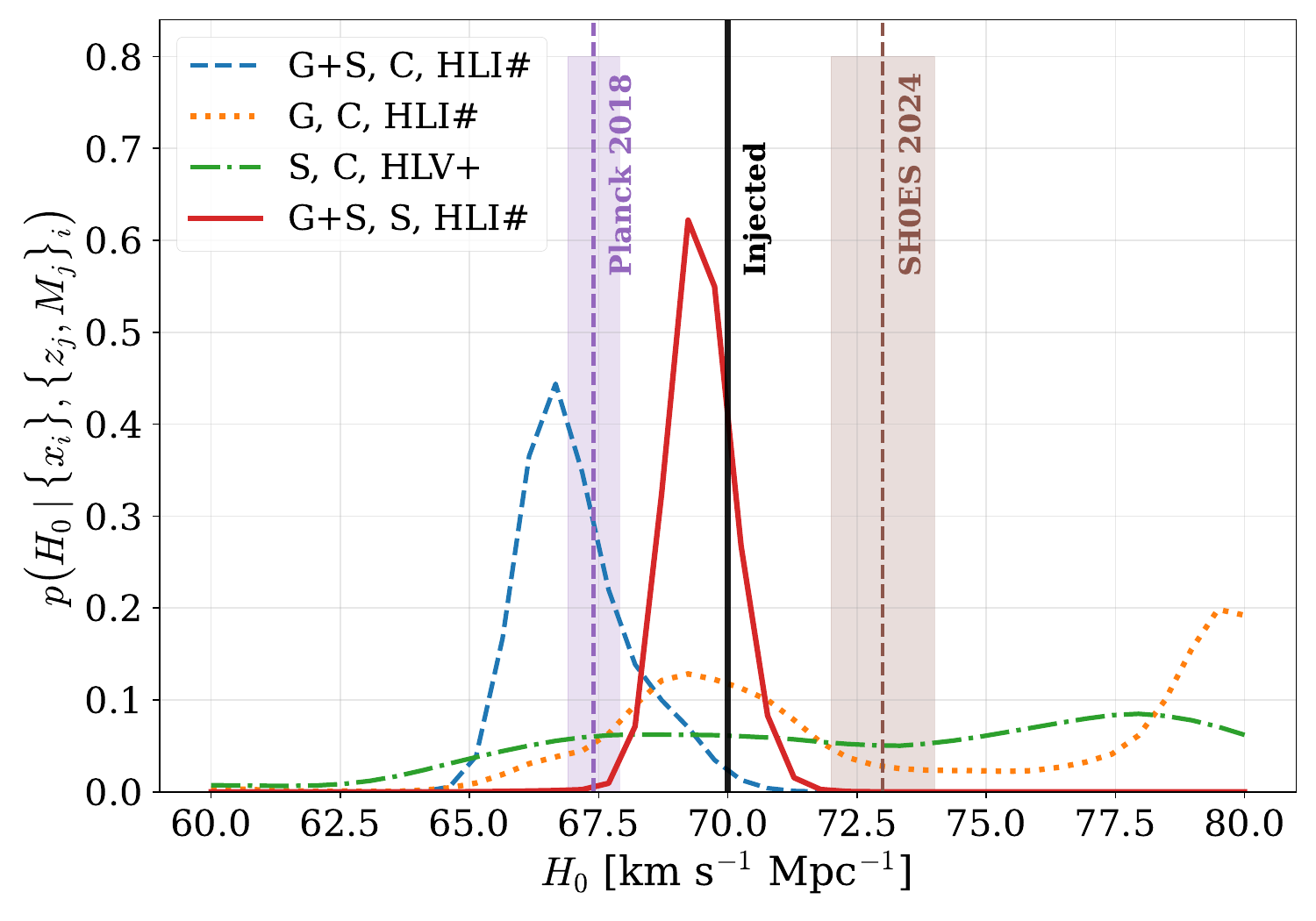}
    
    \caption{$H_0$ posterior predicted to be obtained from the accumulation of a year's worth of golden and silver dark siren follow-up observations with $d_L\leq 980$~MPC under each future network. The first labels G+S, G, and S denote golden and silver dark sirens, only golden dark sirens, and only silver dark sirens respectively. The second labels C and S denote either COSMOS field or SHELA field as mock sky background. The last labels HLI\#, HLV$+$ denote each network configuration. }
    \label{fig:H0posterior}
\end{figure}

Figure~\ref{fig:H0posterior} illustrates the $H_0$ posteriors resulting from each mock data challenge for the different GW networks, HETDEX catalogs and the number of years of GW observations. We did not apply any luminosity weighting in our analysis as our mock universe assumes no correlation between the probability of a galaxy hosting a CBC merger and its luminosity. However, for this study, we restricted the luminosity distance of simulated events to be smaller than $980$~Mpc (i.e., $z\leq0.2$ with $H_0=70\,\mathrm{km\,s^{-1}\,Mpc^{-1}}$) and the absolute magnitudes of the true hosts to be greater than $M_g=-17.6$. These choices ensure that our magnitude-completeness assumption remains valid.  A brief summary of the results of our analysis is as follows:
\begin{itemize}
    \item HLI\# (silver + golden; 25 events; COSMOS field): $H_0 = 66.35^{+1.10\,(+1.7\%)}_{-0.78\,(-1.2\%)}\ \mathrm{km\,s^{-1}\,Mpc^{-1}}$
    \item HLI\# (golden only; 3 events; COSMOS field): $H_0 = 70.14^{+8.96\,(+12.8\%)}_{-2.76\,(-3.9\%)}\ \mathrm{km\,s^{-1}\,Mpc^{-1}}$
    \item HLV$+$ (silver; 4 events; COSMOS field): $H_0 = 67.81^{+8.83\,(+13.0\%)}_{-6.06\,(-8.9\%)}\ \mathrm{km\,s^{-1}\,Mpc^{-1}}$
    \item HLI\# (silver + golden; 25 events; SHELA field): $H_0 = 69.19^{+0.67\,(+1.0\%)}_{-0.61\,(-0.9\%)}\ \mathrm{km\,s^{-1}\,Mpc^{-1}}$.
\end{itemize}
 \begin{figure*}[bt!]
    \centering
    \includegraphics[width=0.48\linewidth]{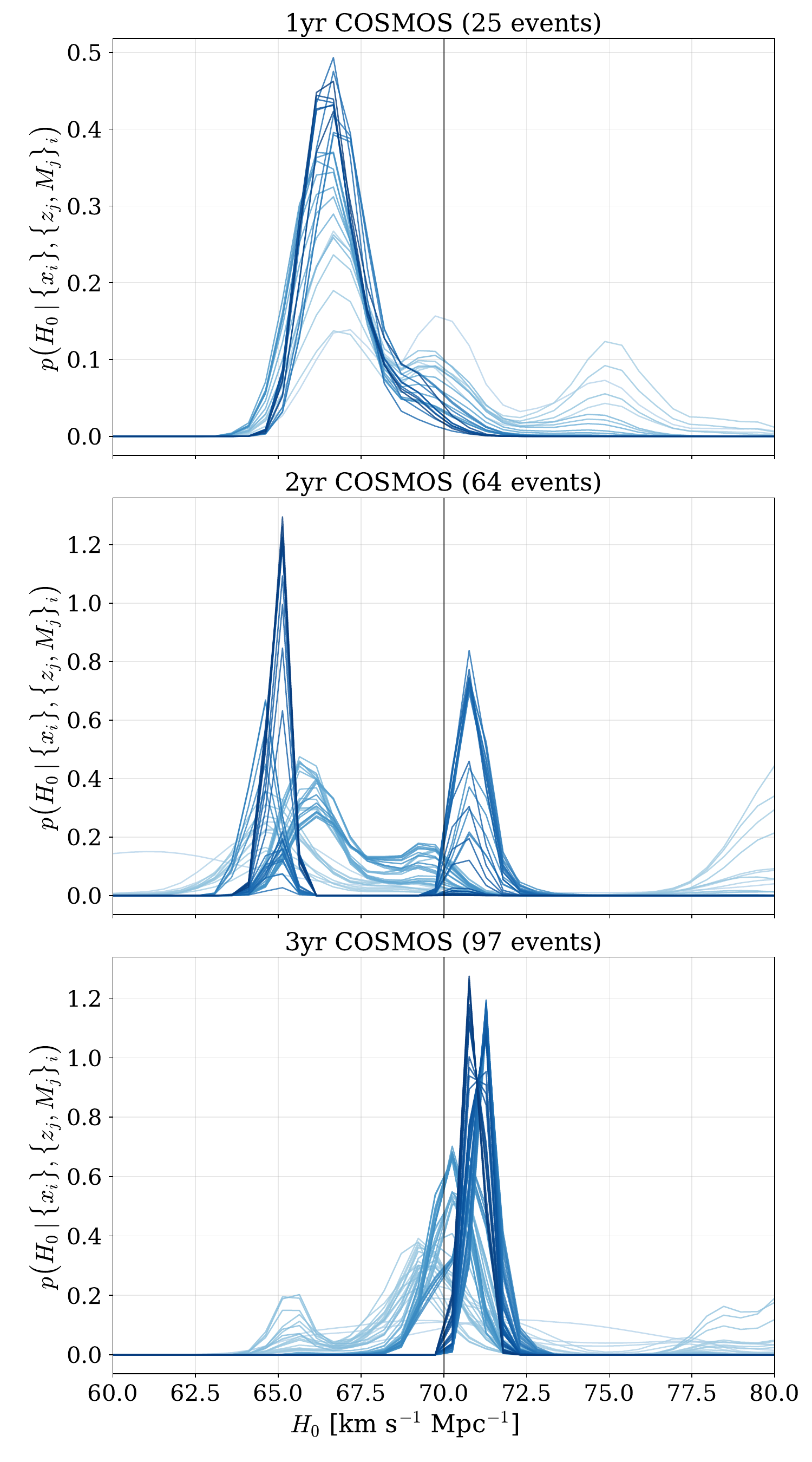}
    \includegraphics[width=0.48\linewidth]{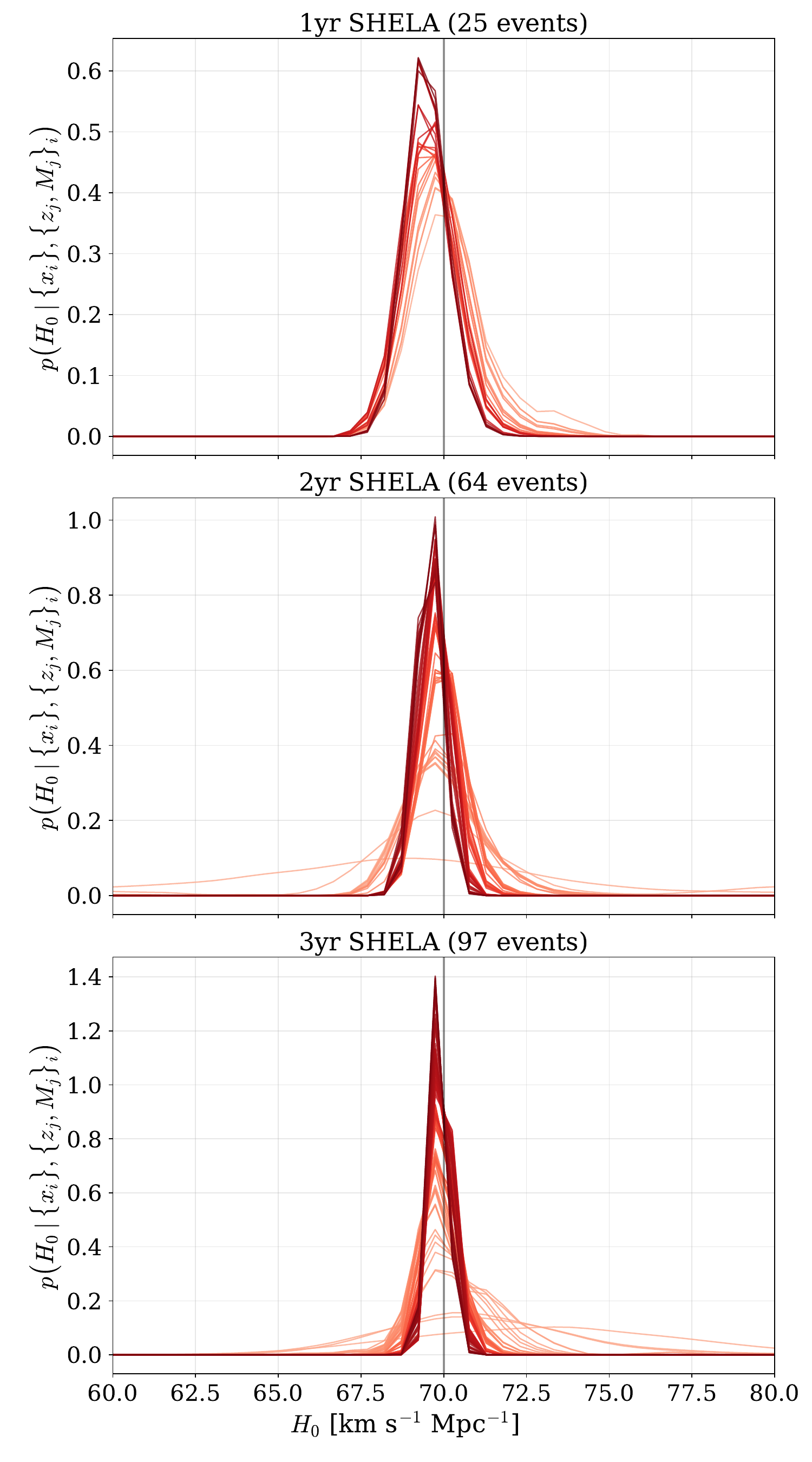}
    \caption{Evolution of the $H_0$ posterior obtained from one, two, and three years of golden and silver dark sirens within $d_L \leq 980$~MPC, using the COSMOS (Left) or SHELA (Right) field as the mock sky background under the HLI\# network. The color progressively darkens as additional events are added to the catalog. The injected value is $H_0 = 70\,\mathrm{km\,s^{-1}\,Mpc^{-1}}$, and no luminosity weighting is applied.}
    \label{fig:evolution}
\end{figure*}

Our results highlight both the strengths and limitations of our mock data challenge. On the positive side, our approach shows that just a handful of golden and silver sirens are enough to place tight constraints on $H_0$.  For example, in the HLI\# configuration, all it takes is 25 events to achieve a $H_0$ measurement that is good to a few percent. However, the results also show that when we limit our mock data challenge to a single region of space -- the COSMOS field -- the posteriors undergo much larger statistical fluctuations due to the field's large-scale structure. In Figure\,~\ref{fig:evolution}, we illustrate this by showing 1-year, 2-year, and 3-year results based on the galaxies of COSMOS and SHELA\null. COSMOS contains a prominent galaxy overdensity at $z\sim0.13$, and a less pronounced one at $z\sim0.18$, violating the assumption of cosmological homogeneity. Because our simulated events span a range of redshifts, they are affected by the large number of possible hosts in these structures differently, leading to persistent fluctuations as additional events are accumulated. The large statistical fluctuation stems from our specific mock simulation procedure. GW events were injected following a Madau-Dickinson SFR history rather than being assigned directly to galaxies within the catalog. This choice was made to keep the GW population modeling independent of the specific galaxy realization used in the mock challenge. For a realistic observational campaign where GW sources are physically coupled to their host environments, LSS typically stabilizes the $H_0$ estimate by providing a consistent environmental prior \citep{Kalomenopoulos:2025qpt, VanWyngarden:2025ogy}. In contrast, the larger area of the SHELA field mitigates the systematic error introduced by cosmic variance, resulting in an unbiased inference that remains stable as the number of events increases. However due to its lower fill-factor, the uncertainty in $H_0$ derived from its analysis is certainly underestimated.

In a realistic scenario, each GW event will occur against a different galaxy field with a different redshift-distribution of potential host galaxies.  Consequently,  while the posterior of any single GW event may be affected by a field's large-scale structure, the statistical fluctuation from an ensemble of events distributed over different fields will average out.  This underscore the importance of accumulating golden and silver dark sirens across multiple, widely separated sky regions to help mitigate the effect of field-specific clustering.

\begin{figure*}
    \centering
    \includegraphics[width=\linewidth]{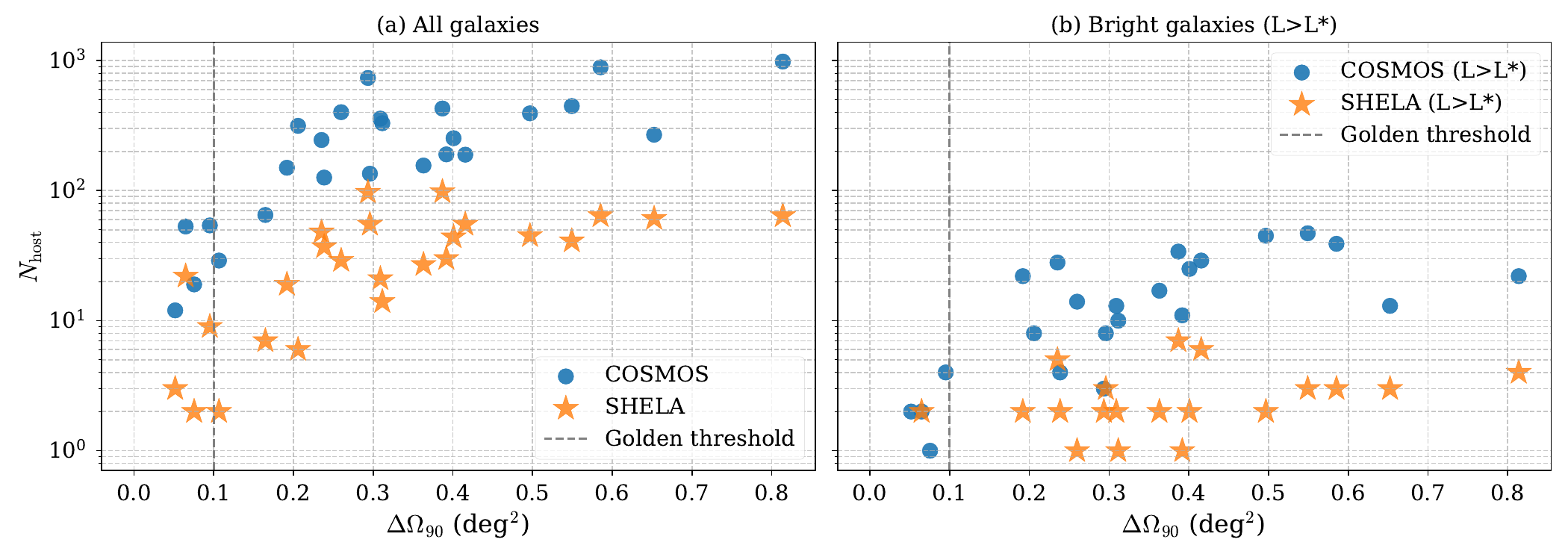}
    \caption{(a) Number of galaxies contained within the 90\% credible sky areas ($\Delta \Omega_{90}$) for each GW injection.  
    (b) Same as panel (a), but showing only galaxies with luminosities exceeding the characteristic luminosity $L^\ast$ in \citep{SDSS:2002vxn}.  
    Blue dots represent simulations using the COSMOS field as the background galaxy catalog, while orange stars represent those using the SHELA field.  
    The vertical dashed line marks the ``golden'' threshold at $\Delta \Omega_{90}=0.1~\mathrm{deg}^2$.  
    The SHELA field, being less complete, typically yields fewer host candidates per event because of its sparser sampling and lower source density.}
    \label{fig:area_selected_galaxy}
\end{figure*}
Figure~\ref{fig:area_selected_galaxy} illustrates the relationship between the 90\% credible sky area ($\Delta \Omega_{90}$) and the number of galaxies enclosed within the corresponding credible volume for each event. In panel~(a), we consider all galaxies within the localization region. As expected, events with smaller sky localization areas contain significantly fewer candidate hosts. The COSMOS field is more densely populated than the SHELA field, so the number of potential hosts is larger, despite having the same GW sky area. This feature also explains the narrower $H_0$ posterior obtained with SHELA (see~Fig.\,~\ref{fig:H0posterior}). 

In our analysis, we begin with a flat prior that every galaxy, regardless of luminosity, has the same likelihood of being the host of a GW event.  Thus, as illustrated by the luminosity function of Figure~\ref{fig:LF}, most of the galaxies contained in the 90\% credible region are faint.  If we were to restrict our analysis to $L > L^\ast$ galaxies, where $L^\ast$ corresponds to $M^\ast=-19.39$ \citep{SDSS:2002vxn}, then the number of possible hosts would drop significantly.  As illustrated in panel~(b) of Figure~\ref{fig:area_selected_galaxy}, events we found meeting the golden threshold ($\Delta \Omega_{90} \leq 0.1\,\mathrm{deg}^2$) have typically fewer than two potential hosts. Furthermore, the majority of silver events are found with fewer than 20 candidates, even with the COSMOS field as the sky background. These results provide an empirical basis for our observational criterion for golden and silver dark sirens. This also suggests that if luminosity weighting were to be used, golden and silver dark sirens with only one or several host candidates dominating the $H_0$ posterior would become much more common. Recent studies by \cite{VanWyngarden:2025ogy} and \cite{Kalomenopoulos:2025qpt} have demonstrated that catalogs with lower completeness can still yield competitive constraints on $H_0$ by incorporating luminosity weighting and galaxy clustering.

\section{Discussion and Conclusions}
\label{sec:conclusion}

The results presented in Section~\ref{sec:results} demonstrate the significant improvement in sky localization and cosmological inference achievable with upgraded gravitational-wave detector networks operating at A$+$ and A$^\#$ sensitivities. As summarized in Table~\ref{tab:expected_number}, the transition from HLV+ to HLI\# increases the number of well-localized dark sirens by more than an order of magnitude. In particular, the HLI\# network produces several golden dark siren events ($\Delta \Omega_{90} \leq 0.1~\mathrm{deg}^2$), marking the first time such precision becomes feasible for binary black hole mergers. The number of silver dark sirens predicted per year in this work is in good agreement with the estimates provided by \citet{Pandey:2024mlo}.

The examples shown in Figs.~\ref{fig:silver_inj} and~\ref{fig:golden_inj} illustrate how improved detector sensitivity directly translates into tighter sky areas and reduced host-galaxy degeneracy. While golden dark sirens may still depend on the luck of not falling in an overdense region of sky, silver sirens should be common and provide a robust statistical path toward an independent measurement of the Hubble constant. 

Combining one year of simulated detections can potentially yield an $H_0$ precision at the percent level for the most sensitive network HLI\#. However it should be noted that the HET can only access about 57\% of the sky, and without an instrument equivalent to VIRUS in the southern hemisphere, it might require more years of observation to achieve the predicted precision. It is worth noting, however, that upcoming wide-field spectroscopic surveys in the south, such as 4MOST on the VISTA telescope \citep{4most}, could serve as a vital supplement to the HET-like program proposed here. Overall, these findings confirm that with improved detector sensitivity and adequate galaxy surveys, dark sirens will eventually become a competitive and independent probe of the local expansion rate of the Universe.

Several limitations in our current analysis should be noted. First, we have not accounted for the contribution of peculiar velocities to the observed redshift, which can introduce an additional uncertainty of $\sim 1$--$2\%$ in $H_0$ for nearby sources. This effect becomes important when the cosmological redshift is comparable to the typical peculiar-velocity dispersion. Second, the galaxy catalogs used in this work cover only small sky areas, making our measurements highly sensitive to the large-scale structure in the host field. A comparison of COSMOS and SHELA shows that field-specific clustering can introduce strong statistical fluctuations that persist with increasing numbers of events. In COSMOS, the limited field size causes introduces a systematic in the $H_0$ posterior that reflects the region's large-scale structure; in SHELA, which is almost 30 times larger, the effect is must less pronounced, though the lower fill-factor artificially narrows the posterior. We expect that for real GW detections, which occur all over the sky, the effects of field-specific clustering will be minimal. However, for mock studies such as that performed here, the effect can dominate the results. Several improvements can be made in the future: (a) computing the selection function individually for each mini-catalog built for golden or silver dark siren (b) performing efficient galaxy clustering algorithm so the marginalization is carried out over independent host environments rather than over many correlated galaxies in the same large-scale structure, and (c) ensuring follow-up catalogs have fill factors near unity, so host-galaxy incompleteness does not introduce field-dependent selection biases. 


\section*{Acknowledgments}

The authors wish to thank Archisman Ghosh, Olivia Curtis, and Craig Wheeler for a careful reading of the manuscript and invaluable comments that helped improve the presentation. Y.D. thanks Yi Qiu, Alberto Salvarese, Giovanni Benetti, Kuan Wang, and Koustav Chandra for helpful discussions. Y.D. is partially supported by NSF grant AST-2307147 and B.S.S. acknowledges support from National Science Foundation (NSF) grants AST-2307147, PHY-2308886 and PHY-2309064.  I.G. acknowledges support from the Network for Neutrinos, Nuclear Astrophysics, and Symmetries (N3AS) Collaboration, NSF grant: PHY-2020275. R.C. and C.G. acknowledges support from the NSF under grant AST-2408358. H.Y.C. is supported by the NSF under Grant PHY-2308752. R.G. is suported by STFC grant ST/V005634/1. D.J. was supported by the NSF under Grant AST-2307026. S.S. acknowledges support from the National Science Foundation under grants NSF-2219212 and NSF-2511145. The authors would also like to acknowledge the LIGO Laboratory computing resources supported by NSF grants:PHY-0757058 and PHY-0823459, and the Gwave cluster maintained by the Institute for Computational and Data Sciences at Penn State University, supported by NSF grants: OAC2346596, OAC2201445, OAC2103662, OAC2018299, and PHY-2110594. The Institute for Gravitation and the Cosmos is supported by the Eberly College of Science and the Office of the Senior Vice President for Research at the Pennsylvania State University. 

HETDEX is led by the University of Texas at Austin McDonald Observatory and Department of Astronomy with participation from the Ludwig-Maximilians-Universität München, Max-Planck-Institut f\"ur Extraterrestrische Physik (MPE), Leibniz-Institut f\"ur Astrophysik Potsdam (AIP), Texas A\&M University, Pennsylvania State University, Institut f\"ur Astrophysik G\"ottingen, The University of Oxford, Max-Planck-Institut f\"ur Astrophysik (MPA), The University of Tokyo and Missouri University of Science and Technology.

Observations for HETDEX were obtained with the Hobby-Eberly Telescope (HET), which is a joint project of the University of Texas at Austin, the Pennsylvania State University, Ludwig-Maximilians-Universität München, and Georg-August-Universit\"at G\"ottingen. The HET is named in honor of its principal benefactors, William P. Hobby and Robert E. Eberly. The Visible Integral-field Replicable Unit Spectrograph (VIRUS) was used for HETDEX observations. VIRUS is a joint project of the University of Texas at Austin, Leibniz-Institut f\"ur Astrophysik Potsdam (AIP), Texas A\&M University, Max-Planck-Institut f\"ur Extraterrestrische Physik (MPE), Ludwig-Maximilians-Universität München, Pennsylvania State University, Institut für Astrophysik Göttingen, University of Oxford, and the Max-Planck-Institut f\"ur Astrophysik (MPA).

The authors acknowledge the Texas Advanced Computing Center (TACC) at The University of Texas at Austin for providing high performance computing, visualization, and storage resources that have contributed to the research results reported within this paper. URL: http://www.tacc.utexas.edu

Funding for HETDEX has been provided by the partner institutions, the National Science Foundation, the State of Texas, the US Air Force, and by generous support from private individuals and foundations.

GAMA is a joint European-Australasian project based around a spectroscopic campaign using the Anglo-Australian Telescope. The GAMA input catalogue is based on data taken from the Sloan Digital Sky Survey and the UKIRT Infrared Deep Sky Survey. Complementary imaging of the GAMA regions was obtained by a number of independent survey programmes including GALEX MIS, VST KiDS, VISTA VIKING, WISE, Herschel-ATLAS, GMRT and ASKAP providing UV to radio coverage. GAMA is funded by the STFC (UK), the ARC (Australia), the AAO, and the participating institutions. The GAMA website is https://www.gama-survey.org/

 \textit{Software:} \texttt{astropy} \citep{astropy2018}, \texttt{GWBENCH} \citep{Borhanian:2020ypi}, \texttt{bilby} \citep{bilby_paper}, \texttt{Matplotlib} \citep{2007matplotlib}, \texttt{NumPy} \citep{Harris_2020}, \texttt{SciPy} \citep{Virtanen_2020}, \texttt{EliXer} \citep{davis_hetdex_2023}.

\bibliography{sample701}

@ARTICLE{Borhanian:2020ypi,
       author = {{Borhanian}, S.},
        title = "{GWBENCH: a novel Fisher information package for gravitational-wave benchmarking}",
      journal = {Classical and Quantum Gravity},
         year = 2021,
        month = aug,
       volume = {38},
       number = {17},
          eid = {175014},
        pages = {175014},
          doi = {10.1088/1361-6382/ac1618},
archivePrefix = {arXiv},
       eprint = {2010.15202},
 primaryClass = {gr-qc},
       adsurl = {https://ui.adsabs.harvard.edu/abs/2021CQGra..38q5014B}
}

@article{Garcia-Quiros:2020qpx,
    author = "Garc{\'\i}a-Quir{\'o}s, Cecilio and Colleoni, Marta and Husa, Sascha and Estell{\'e}s, H{\'e}ctor and Pratten, Geraint and Ramos-Buades, Antoni and Mateu-Lucena, Maite and Jaume, Rafel",
    title = "{Multimode frequency-domain model for the gravitational wave signal from nonprecessing black-hole binaries}",
    eprint = "2001.10914",
    archivePrefix = "arXiv",
    primaryClass = "gr-qc",
    doi = "10.1103/PhysRevD.102.064002",
    journal = {\prd},
    volume = "102",
    number = "6",
    pages = "064002",
    year = "2020"
}

@article{LIGOScientific:2020kqk,
    author = "Abbott, R. and others",
    collaboration = "LIGO Scientific, Virgo",
    title = "{Population Properties of Compact Objects from the Second LIGO-Virgo Gravitational-Wave Transient Catalog}",
    eprint = "2010.14533",
    archivePrefix = "arXiv",
    primaryClass = "astro-ph.HE",
    reportNumber = "LIGO-P2000077",
    doi = "10.3847/2041-8213/abe949",
    journal = {\apjl},
    volume = "913",
    number = "1",
    pages = "L7",
    year = "2021"
}

@article{Wysocki:2018mpo,
    author = "Wysocki, Daniel and Lange, Jacob and O'Shaughnessy, Richard",
    title = "{Reconstructing phenomenological distributions of compact binaries via gravitational wave observations}",
    eprint = "1805.06442",
    archivePrefix = "arXiv",
    primaryClass = "gr-qc",
    reportNumber = "LIGO P1800107, LIGO-P1800107",
    doi = "10.1103/PhysRevD.100.043012",
    journal = {\prd},
    volume = "100",
    number = "4",
    pages = "043012",
    year = "2019"
}

@article{bilby_paper,
    author = "Ashton, Gregory and others",
    title = "{BILBY: A user-friendly Bayesian inference library for gravitational-wave astronomy}",
    eprint = "1811.02042",
    archivePrefix = "arXiv",
    primaryClass = "astro-ph.IM",
    doi = "10.3847/1538-4365/ab06fc",
    journal = {\apjs},
    volume = "241",
    number = "2",
    pages = "27",
    year = "2019"
}

@unpublished{relbin_bilby,
    author = "Krishna, Kruthi and Vijaykumar, Aditya and Ganguly, Apratim and Talbot, Colm and Biscoveanu, Sylvia and George, Richard N. and Williams, Natalie and Zimmerman, Aaron",
    title = "{Accelerated parameter estimation in Bilby with relative binning}",
    eprint = "2312.06009",
    archivePrefix = "arXiv",
    primaryClass = "gr-qc",
    month = "12",
    year = "2023",
    note = {\url{https://arxiv.org/pdf/2312.06009}}
}

@unpublished{relbin_cornish,
    author = "Cornish, Neil J.",
    title = "{Fast Fisher Matrices and Lazy Likelihoods}",
    eprint = "1007.4820",
    archivePrefix = "arXiv",
    primaryClass = "gr-qc",
    month = "7",
    year = "2010",
    note = {\url{https://arxiv.org/pdf/1007.4820}}
}

@article{KAGRA:2021duu,
    author = "Abbott, R. and others",
    collaboration = "KAGRA, VIRGO, LIGO Scientific",
    title = "{Population of Merging Compact Binaries Inferred Using Gravitational Waves through GWTC-3}",
    eprint = "2111.03634",
    archivePrefix = "arXiv",
    primaryClass = "astro-ph.HE",
    reportNumber = "LIGO-P2100239 ; Data release: https://zenodo.org/record/5655785, LIGO-P2100239",
    doi = "10.1103/PhysRevX.13.011048",
    journal = "Phys. Rev. X",
    volume = "13",
    number = "1",
    pages = "011048",
    year = "2023"
}

@article{LIGOScientific:2018jsj,
	title        = {{Binary Black Hole Population Properties Inferred from the First and Second Observing Runs of Advanced LIGO and Advanced Virgo}},
	author       = {Abbott, B. P. and others},
	year         = 2019,
	journal      = {\apjl},
	volume       = 882,
	number       = 2,
	pages        = {L24},
	doi          = {10.3847/2041-8213/ab3800},
	collaboration = {LIGO Scientific, Virgo},
	eprint       = {1811.12940},
	archiveprefix = {arXiv},
	primaryclass = {astro-ph.HE},
	reportnumber = {LIGO-P1800324}
}

@unpublished{Zackay:2018qdy,
    author = "Zackay, Barak and Dai, Liang and Venumadhav, Tejaswi",
    title = "{Relative Binning and Fast Likelihood Evaluation for Gravitational Wave Parameter Estimation}",
    eprint = "1806.08792",
    archivePrefix = "arXiv",
    primaryClass = "astro-ph.IM",
    month = "6",
    year = "2018",
    note = {https://arxiv.org/abs/1806.08792}
}

@article{Scoville:2006vq,
    author = "Scoville, Nick and others",
    title = "{The Cosmic Evolution Survey (COSMOS): Overview}",
    eprint = "astro-ph/0612305",
    archivePrefix = "arXiv",
    doi = "10.1086/516585",
    journal = {\apjs},
    volume = "172",
    pages = "1--8",
    year = "2007"
}

@article{Casey:2022amu,
    author = "Casey, Caitlin M. and others",
    title = "{COSMOS-Web: An Overview of the JWST Cosmic Origins Survey}",
    eprint = "2211.07865",
    archivePrefix = "arXiv",
    primaryClass = "astro-ph.GA",
    doi = "10.3847/1538-4357/acc2bc",
    journal = {\apj},
    volume = "954",
    number = "1",
    pages = "31",
    year = "2023"
}

@ARTICLE{Hill2021,
       author = {{Hill}, Gary J. and {Lee}, Hanshin and {MacQueen}, Phillip J. and {Kelz}, Andreas and {Drory}, Niv and {Vattiat}, Brian L. and {Good}, John M. and {Ramsey}, Jason and {Kriel}, Herman and {Peterson}, Trent and {DePoy}, D.~L. and {Gebhardt}, Karl and {Marshall}, J.~L. and {Tuttle}, Sarah E. and {Bauer}, Svend M. and {Chonis}, Taylor S. and {Fabricius}, Maximilian H. and {Froning}, Cynthia and {H{\"a}user}, Marco and {Indahl}, Briana L. and {Jahn}, Thomas and {Landriau}, Martin and {Leck}, Ron and {Montesano}, Francesco and {Prochaska}, Travis and {Snigula}, Jan M. and {Zeimann}, Greg and {Bryant}, Randy and {Damm}, George and {Fowler}, J.~R. and {Janowiecki}, Steven and {Martin}, Jerry and {Mrozinski}, Emily and {Odewahn}, Stephen and {Rostopchin}, Sergey and {Shetrone}, Matthew and {Spencer}, Renny and {Mentuch Cooper}, Erin and {Armandroff}, Taft and {Bender}, Ralf and {Dalton}, Gavin and {Hopp}, Ulrich and {Komatsu}, Eiichiro and {Nicklas}, Harald and {Ramsey}, Lawrence W. and {Roth}, Martin M. and {Schneider}, Donald P. and {Sneden}, Chris and {Steinmetz}, Matthias},
        title = "{The HETDEX Instrumentation: Hobby-Eberly Telescope Wide-field Upgrade and VIRUS}",
      journal = {\aj},
         year = 2021,
        month = dec,
       volume = {162},
       number = {6},
          eid = {298},
        pages = {298},
          doi = {10.3847/1538-3881/ac2c02},
archivePrefix = {arXiv},
       eprint = {2110.03843},
 primaryClass = {astro-ph.IM},
       adsurl = {https://ui.adsabs.harvard.edu/abs/2021AJ....162..298H}
}

@INPROCEEDINGS{Ramsey1998,
       author = {{Ramsey}, Lawrence W. and {Adams}, M.~T. and {Barnes}, Thomas G. and {Booth}, John A. and {Cornell}, Mark E. and {Fowler}, James R. and {Gaffney}, Niall I. and {Glaspey}, John W. and {Good}, John M. and {Hill}, Gary J. and {Kelton}, Philip W. and {Krabbendam}, Victor L. and {Long}, L. and {MacQueen}, Phillip J. and {Ray}, Frank B. and {Ricklefs}, Randall L. and {Sage}, J. and {Sebring}, Thomas A. and {Spiesman}, W.~J. and {Steiner}, M.},
        title = "{Early performance and present status of the Hobby-Eberly Telescope}",
    booktitle = {Advanced Technology Optical/IR Telescopes VI},
         year = 1998,
       editor = {{Stepp}, Larry M.},
       series = {Society of Photo-Optical Instrumentation Engineers (SPIE) Conference Series},
       volume = {3352},
        month = aug,
        pages = {34-42},
          doi = {10.1117/12.319287},
       adsurl = {https://ui.adsabs.harvard.edu/abs/1998SPIE.3352...34R}
}

@ARTICLE{Zeimann2024,
       author = {{Zeimann}, Gregory R. and {Debski}, Maya H. and {Schneider}, Donald P. and {Bowman}, William P. and {Drory}, Niv and {Hill}, Gary J. and {Lee}, Hanshin and {MacQueen}, Phillip and {Shetrone}, Matthew},
        title = "{The Hobby{\textendash}Eberly Telescope VIRUS Parallel Survey (HETVIPS)}",
      journal = {\apj},
         year = 2024,
        month = may,
       volume = {966},
       number = {1},
          eid = {14},
        pages = {14},
          doi = {10.3847/1538-4357/ad35b8},
archivePrefix = {arXiv},
       eprint = {2405.00585},
 primaryClass = {astro-ph.IM},
       adsurl = {https://ui.adsabs.harvard.edu/abs/2024ApJ...966...14Z}
}

@ARTICLE{Huchra1973,
       author = {{Huchra}, J. and {Sargent}, W.~L.~W.},
        title = "{The space density of the Markarian galaxies including a region of the south galactic hemisphere.}",
      journal = {\apj},
         year = 1973,
        month = dec,
       volume = {186},
        pages = {433-443},
          doi = {10.1086/152510},
       adsurl = {https://ui.adsabs.harvard.edu/abs/1973ApJ...186..433H}
}

@article{Schechter:1976iz,
    author = "Schechter, P.",
    title = "{An analytic expression for the luminosity function for galaxies}",
    doi = "10.1086/154079",
    journal = "Astrophys. J.",
    volume = "203",
    pages = "297--306",
    year = "1976"
}

@ARTICLE{Gallego2002,
       author = {{Gallego}, J. and {Garc{\'\i}a-Dab{\'o}}, C.~E. and {Zamorano}, J. and {Arag{\'o}n-Salamanca}, A. and {Rego}, M.},
        title = "{The [O II] {\ensuremath{\lambda}}3727 Luminosity Function of the Local Universe}",
      journal = {\apjl},
         year = 2002,
        month = may,
       volume = {570},
       number = {1},
        pages = {L1-L4},
          doi = {10.1086/340830},
archivePrefix = {arXiv},
       eprint = {astro-ph/0210062},
 primaryClass = {astro-ph},
       adsurl = {https://ui.adsabs.harvard.edu/abs/2002ApJ...570L...1G}
}

@article{Madau:1996hu,
    author = "Madau, Piero",
    editor = "Holt, Stephen S. and Mundy, Lee G.",
    title = "{Cosmic star formation history}",
    eprint = "astro-ph/9612157",
    archivePrefix = "arXiv",
    doi = "10.1063/1.52821",
    journal = "AIP Conf. Proc.",
    volume = "393",
    number = "1",
    pages = "481",
    year = "1997"
}

@ARTICLE{Salzer2000,
       author = {{Salzer}, John J. and {Gronwall}, Caryl and {Lipovetsky}, Valentin A. and {Kniazev}, Alexei and {Moody}, J. Ward and {Boroson}, Todd A. and {Thuan}, Trinh X. and {Izotov}, Yuri I. and {Herrero}, Jose L. and {Frattare}, Lisa M.},
        title = "{The KPNO International Spectroscopic Survey. I. Description of the Survey}",
      journal = {\aj},
         year = 2000,
        month = jul,
       volume = {120},
       number = {1},
        pages = {80-94},
          doi = {10.1086/301418},
archivePrefix = {arXiv},
       eprint = {astro-ph/0004074},
 primaryClass = {astro-ph},
       adsurl = {https://ui.adsabs.harvard.edu/abs/2000AJ....120...80S}
}

@ARTICLE{Ciardullo2013,
       author = {{Ciardullo}, Robin and {Gronwall}, Caryl and {Adams}, Joshua J. and {Blanc}, Guillermo A. and {Gebhardt}, Karl and {Finkelstein}, Steven L. and {Jogee}, Shardha and {Hill}, Gary J. and {Drory}, Niv and {Hopp}, Ulrich and {Schneider}, Donald P. and {Zeimann}, Gregory R. and {Dalton}, Gavin B.},
        title = "{The HETDEX Pilot Survey. IV. The Evolution of [O II] Emitting Galaxies from z \raisebox{-0.5ex}\textasciitilde 0.5 to z \raisebox{-0.5ex}\textasciitilde 0}",
      journal = {\apj},
         year = 2013,
        month = may,
       volume = {769},
       number = {1},
          eid = {83},
        pages = {83},
          doi = {10.1088/0004-637X/769/1/83},
archivePrefix = {arXiv},
       eprint = {1304.5537},
 primaryClass = {astro-ph.CO},
       adsurl = {https://ui.adsabs.harvard.edu/abs/2013ApJ...769...83C}
}

@ARTICLE{Hawkins2021,
       author = {{Hawkins}, Keith and {Zeimann}, Greg and {Sneden}, Chris and {Cooper}, Erin Mentuch and {Gebhardt}, Karl and {Bond}, Howard E. and {Carrillo}, Andreia and {Casey}, Caitlin M. and {Castanheira}, Barbara G. and {Ciardullo}, Robin and {Davis}, Dustin and {Farrow}, Daniel J. and {Finkelstein}, Steven L. and {Hill}, Gary J. and {Kelz}, Andreas and {Liu}, Chenxu and {Shetrone}, Matthew and {Schneider}, Donald P. and {Starkenburg}, Else and {Steinmetz}, Matthias and {Wheeler}, J. Craig and {Hetdex Collaboration}},
        title = "{The Stars of the HETDEX Survey. I. Radial Velocities and Metal-poor Stars from Low-resolution Stellar Spectra}",
      journal = {\apj},
         year = 2021,
        month = apr,
       volume = {911},
       number = {2},
          eid = {108},
        pages = {108},
          doi = {10.3847/1538-4357/abe9bd},
archivePrefix = {arXiv},
       eprint = {2102.06224},
 primaryClass = {astro-ph.GA},
       adsurl = {https://ui.adsabs.harvard.edu/abs/2021ApJ...911..108H}
}

@article{Schmidt:1968kn,
    author = "Schmidt, Maarten",
    title = "{Space Distribution and Luminosity Functions of Quasi-Stellar Radio Sources}",
    doi = "10.1086/149446",
    journal = {\apj},
    volume = "151",
    pages = "393",
    year = "1968"
}

@unpublished{Cousins:2025bas,
    author = "Cousins, Bryce and Schumacher, Kristen and Chung, Adrian Ka-Wai and Talbot, Colm and Callister, Thomas and Holz, Daniel E. and Yunes, Nicol{\'a}s",
    title = "{The Stochastic Siren: Astrophysical Gravitational-Wave Background Measurements of the Hubble Constant}",
    eprint = "2503.01997",
    archivePrefix = "arXiv",
    primaryClass = "astro-ph.CO",
    month = "3",
    year = "2025",
    note = {\url{https://arxiv.org/pdf/2503.01997}}
}

@article{Taylor:2012db,
    author = "Taylor, Stephen R. and Gair, Jonathan R.",
    title = "{Cosmology with the lights off: standard sirens in the Einstein Telescope era}",
    eprint = "1204.6739",
    archivePrefix = "arXiv",
    primaryClass = "astro-ph.CO",
    doi = "10.1103/PhysRevD.86.023502",
    journal = {\prd},
    volume = "86",
    pages = "023502",
    year = "2012"
}

@article{Mastrogiovanni:2021wsd,
    author = "Mastrogiovanni, S. and Leyde, K. and Karathanasis, C. and Chassande-Mottin, E. and Steer, D. A. and Gair, J. and Ghosh, A. and Gray, R. and Mukherjee, S. and Rinaldi, S.",
    title = "{On the importance of source population models for gravitational-wave cosmology}",
    eprint = "2103.14663",
    archivePrefix = "arXiv",
    primaryClass = "gr-qc",
    doi = "10.1103/PhysRevD.104.062009",
    journal = {\prd},
    volume = "104",
    number = "6",
    pages = "062009",
    year = "2021"
}

@article{Namikawa:2015prh,
    author = "Namikawa, Toshiya and Nishizawa, Atsushi and Taruya, Atsushi",
    title = "{Anisotropies of gravitational-wave standard sirens as a new cosmological probe without redshift information}",
    eprint = "1511.04638",
    archivePrefix = "arXiv",
    primaryClass = "astro-ph.CO",
    doi = "10.1103/PhysRevLett.116.121302",
    journal = {\prl},
    volume = "116",
    number = "12",
    pages = "121302",
    year = "2016"
}

@article{Bera:2020jhx,
    author = "Bera, Sayantani and Rana, Divya and More, Surhud and Bose, Sukanta",
    title = "{Incompleteness Matters Not: Inference of $H_0$ from Binary Black Hole{\textendash}Galaxy Cross-correlations}",
    eprint = "2007.04271",
    archivePrefix = "arXiv",
    primaryClass = "astro-ph.CO",
    reportNumber = "LIGO-P2000239-v2",
    doi = "10.3847/1538-4357/abb4e0",
    journal = {\apj},
    volume = "902",
    number = "1",
    pages = "79",
    year = "2020"
}

@article{Mukherjee:2022afz,
    author = "Mukherjee, Suvodip and Krolewski, Alex and Wandelt, Benjamin D. and Silk, Joseph",
    title = "{Cross-correlating dark sirens and galaxies: constraints on $H_0$ from GWTC-3 of LIGO-Virgo-KAGRA}",
    eprint = "2203.03643",
    archivePrefix = "arXiv",
    primaryClass = "astro-ph.CO",
    doi = "10.3847/1538-4357/ad7d90",
    journal = {\apj},
    volume = "975",
    number = "2",
    pages = "189",
    year = "2024"
}

@article{Borhanian:2020vyr,
    author = "Borhanian, Ssohrab and Dhani, Arnab and Gupta, Anuradha and Arun, K. G. and Sathyaprakash, B. S.",
    title = "{Dark Sirens to Resolve the Hubble{\textendash}Lema{\^\i}tre Tension}",
    eprint = "2007.02883",
    archivePrefix = "arXiv",
    primaryClass = "astro-ph.CO",
    reportNumber = "LIGO document number LIGO-P2000229",
    doi = "10.3847/2041-8213/abcaf5",
    journal = {\apjl},
    volume = "905",
    number = "2",
    pages = "L28",
    year = "2020"
}

@misc{Aplus,
  author       = {Barsotti,Lisa and Fritschel, Peter and Evans, Matthew and  Gras, Slawomir},
  title        = {Updated Advanced LIGO sensitivity design curve, Technical Notes, LIGO-T1800044},
  year         = {2018},
  url          = {https://dcc.ligo.org/LIGO-T1800044/public},
  note         = {Accessed: 2025-10-15},
  urldate      = {2025-10-15},
  howpublished = {\url{https://dcc.ligo.org/LIGO-T1800044/public}}
}

@article{Branchesi:2023mws,
    author = "Branchesi, Marica and others",
    title = "{Science with the Einstein Telescope: a comparison of different designs}",
    eprint = "2303.15923",
    archivePrefix = "arXiv",
    primaryClass = "gr-qc",
    reportNumber = "ET-0084A-23",
    doi = "10.1088/1475-7516/2023/07/068",
    journal = {\jcap},
    volume = "07",
    pages = "068",
    year = "2023"
}

@article{Bailes:2021tot,
    author = "Bailes, M. and others",
    title = "{Gravitational-wave physics and astronomy in the 2020s and 2030s}",
    doi = "10.1038/s42254-021-00303-8",
    journal = "Nature Rev. Phys.",
    volume = "3",
    number = "5",
    pages = "344--366",
    year = "2021"
}

@ARTICLE{moore2025desidr2galaxyluminosity,
       author = {{Moore}, Samuel G. and {Cole}, Shaun and {Wilson}, Michael and {Norberg}, Peder and {Moustakas}, John and {Aguilar}, J. and {Ahlen}, S. and {Anand}, A. and {Bianchi}, D. and {Brooks}, D. and {Castander}, F.~J. and {Claybaugh}, T. and {Cuceu}, A. and {de la Macorra}, A. and {Dey}, Arjun and {Dey}, Biprateep and {Ferraro}, S. and {Font-Ribera}, A. and {Forero-Romero}, J.~E. and {Gaztanaga}, E. and {Gontcho}, S. Gontcho A and {Gutierrez}, G. and {Herrera-Alcantar}, H.~K. and {Honscheid}, K. and {Ishak}, M. and {Joyce}, R. and {Juneau}, S. and {Kehoe}, R. and {Kisner}, T. and {Koposov}, S.~E. and {Kremin}, A. and {Lahav}, O. and {Lamman}, C. and {Landriau}, M. and {Le Guillou}, L. and {Levi}, M.~E. and {Manera}, M. and {Meisner}, A. and {Miquel}, R. and {Nadathur}, S. and {Percival}, W.~J. and {Poppett}, C. and {Prada}, F. and {Ross}, A.~J. and {Rossi}, G. and {Sanchez}, E. and {Schlegel}, D. and {Schubnell}, M. and {Seo}, H. and {Silber}, J. and {Sprayberry}, D. and {Tarl{\'e}}, G. and {Weaver}, B.~A. and {Wechsler}, R.~H. and {Zhou}, R. and {Zou}, H.},
        title = "{DESI DR2 Galaxy Luminosity Functions}",
      journal = {arXiv e-prints},
         year = 2025,
        month = nov,
          eid = {arXiv:2511.01803},
        pages = {arXiv:2511.01803},
          doi = {10.48550/arXiv.2511.01803},
archivePrefix = {arXiv},
       eprint = {2511.01803},
 primaryClass = {astro-ph.GA},
       adsurl = {https://ui.adsabs.harvard.edu/abs/2025arXiv251101803M}
}

@unpublished{LIGOScientific:2025pvj,
    author = "Abac, A. G. and others",
    collaboration = "LIGO Scientific, VIRGO, KAGRA",
    title = "{GWTC-4.0: Population Properties of Merging Compact Binaries}",
    eprint = "2508.18083",
    archivePrefix = "arXiv",
    primaryClass = "astro-ph.HE",
    reportNumber = "LIGO-P2400004",
    month = "8",
    year = "2025",
    note = {\url{https://arxiv.org/pdf/2508.18083}}
}

@article{CosmoVerseNetwork:2025alb,
    author = "Di Valentino, Eleonora and others",
    collaboration = "CosmoVerse Network",
    title = "{The CosmoVerse White Paper: Addressing observational tensions in cosmology with systematics and fundamental physics}",
    eprint = "2504.01669",
    archivePrefix = "arXiv",
    primaryClass = "astro-ph.CO",
    doi = "10.1016/j.dark.2025.101965",
    journal = "Phys. Dark Univ.",
    volume = "49",
    pages = "101965",
    year = "2025"
}

@misc{Asharp,
  author       = {Kuns, Kevin and Fritschel, Peter },
  title        = {A\# Strain Sensitivity, Technical Notes, LIGO-T2300041},
  year         = {2023},
  url          = {https://dcc.ligo.org/LIGO-T2300041/public},
  note         = {Accessed: 2025-10-15},
  urldate      = {2025-10-15},
  howpublished = {\url{https://dcc.ligo.org/LIGO-T2300041/public}}
}

@article{Riess:2021jrx,
    author = "Riess, Adam G. and others",
    title = "{A Comprehensive Measurement of the Local Value of the Hubble Constant with 1 km s$^{−1}$ Mpc$^{−1}$ Uncertainty from the Hubble Space Telescope and the SH0ES Team}",
    eprint = "2112.04510",
    archivePrefix = "arXiv",
    primaryClass = "astro-ph.CO",
    doi = "10.3847/2041-8213/ac5c5b",
    journal = {\apjl},
    volume = "934",
    number = "1",
    pages = "L7",
    year = "2022"
}

@ARTICLE{Riess:2025chq,
       author = {{Riess}, Adam G. and {Li}, Siyang and {Anand}, Gagandeep S. and {Yuan}, Wenlong and {Breuval}, Louise and {Casertano}, Stefano and {Macri}, Lucas M. and {Scolnic}, Dan and {Murakami}, Yukei S. and {Filippenko}, Alexei V. and {Brink}, Thomas G.},
        title = "{The Perfect Host: JWST Cepheid Observations in a Background-free Type Ia Supernova Host Confirm No Bias in Hubble-constant Measurements}",
      journal = {\apjl},
         year = 2025,
        month = oct,
       volume = {992},
       number = {2},
          eid = {L34},
        pages = {L34},
          doi = {10.3847/2041-8213/ae0ad6},
archivePrefix = {arXiv},
       eprint = {2509.01667},
 primaryClass = {astro-ph.CO},
       adsurl = {https://ui.adsabs.harvard.edu/abs/2025ApJ...992L..34R}
}

@article{Dalya:2018cnd,
    author = "D{\'a}lya, Gergely and Galg{\'o}czi, G{\'a}bor and Dobos, L{\'a}szl{\'o} and Frei, Zsolt and Heng, Ik Siong and Macas, Ronaldas and Messenger, Christopher and Raffai, P{\'e}ter and de Souza, Rafael S.",
    title = "{GLADE: A galaxy catalogue for multimessenger searches in the advanced gravitational-wave detector era}",
    eprint = "1804.05709",
    archivePrefix = "arXiv",
    primaryClass = "astro-ph.HE",
    doi = "10.1093/mnras/sty1703",
    journal = "Mon. Not. Roy. Astron. Soc.",
    volume = "479",
    number = "2",
    pages = "2374--2381",
    year = "2018"
}

@article{Ezquiaga:2022zkx,
    author = "Ezquiaga, Jose Mar{\'\i}a and Holz, Daniel E.",
    title = "{Spectral Sirens: Cosmology from the Full Mass Distribution of Compact Binaries}",
    eprint = "2202.08240",
    archivePrefix = "arXiv",
    primaryClass = "astro-ph.CO",
    doi = "10.1103/PhysRevLett.129.061102",
    journal = "Phys. Rev. Lett.",
    volume = "129",
    number = "6",
    pages = "061102",
    year = "2022"
}

@article{Dalya:2021ewn,
    author = "D{\'a}lya, G. and others",
    title = "{GLADE+~: an extended galaxy catalogue for multimessenger searches with advanced gravitational-wave detectors}",
    eprint = "2110.06184",
    archivePrefix = "arXiv",
    primaryClass = "astro-ph.CO",
    doi = "10.1093/mnras/stac1443",
    journal = "Mon. Not. Roy. Astron. Soc.",
    volume = "514",
    number = "1",
    pages = "1403--1411",
    year = "2022"
}

@article{Alfradique:2025tbj,
    author = "Alfradique, Viviane and Bom, Cl{\'e}cio R. and Castro, Tiago",
    title = "{Systematic bias in dark siren statistical methods and its impact on Hubble constant measurement}",
    eprint = "2503.18887",
    archivePrefix = "arXiv",
    primaryClass = "astro-ph.CO",
    doi = "10.1103/vd36-3mys",
    journal = {\prd},
    volume = "112",
    number = "6",
    pages = "063561",
    year = "2025"
}

@article{Verde:2019ivm,
    author = "Verde, L. and Treu, T. and Riess, A. G.",
    title = "{Tensions between the Early and the Late Universe}",
    eprint = "1907.10625",
    archivePrefix = "arXiv",
    primaryClass = "astro-ph.CO",
    doi = "10.1038/s41550-019-0902-0",
    journal = "Nature Astron.",
    volume = "3",
    pages = "891",
    year = "2019"
}

@article{Gupta:2022fwd,
    author = "Gupta, Ish",
    title = "{Using grey sirens to resolve the Hubble{\textendash}Lema{\^\i}tre tension}",
    eprint = "2212.00163",
    archivePrefix = "arXiv",
    primaryClass = "gr-qc",
    doi = "10.1093/mnras/stad2115",
    journal = {\mnras},
    volume = "524",
    number = "3",
    pages = "3537--3558",
    year = "2023"
}

@article{Poulin:2018cxd,
    author = "Poulin, Vivian and Smith, Tristan L. and Karwal, Tanvi and Kamionkowski, Marc",
    title = "{Early Dark Energy Can Resolve The Hubble Tension}",
    eprint = "1811.04083",
    archivePrefix = "arXiv",
    primaryClass = "astro-ph.CO",
    doi = "10.1103/PhysRevLett.122.221301",
    journal = {\prl},
    volume = "122",
    number = "22",
    pages = "221301",
    year = "2019"
}

@article{DiValentino:2019qzk,
    author = "Di Valentino, Eleonora and Melchiorri, Alessandro and Silk, Joseph",
    title = "{Planck evidence for a closed Universe and a possible crisis for cosmology}",
    eprint = "1911.02087",
    archivePrefix = "arXiv",
    primaryClass = "astro-ph.CO",
    doi = "10.1038/s41550-019-0906-9",
    journal = "Nature Astron.",
    volume = "4",
    number = "2",
    pages = "196--203",
    year = "2019"
}

@article{DiValentino:2021izs,
    author = "Di Valentino, Eleonora and Mena, Olga and Pan, Supriya and Visinelli, Luca and Yang, Weiqiang and Melchiorri, Alessandro and Mota, David F. and Riess, Adam G. and Silk, Joseph",
    title = "{In the realm of the Hubble tension{\textemdash}a review of solutions}",
    eprint = "2103.01183",
    archivePrefix = "arXiv",
    primaryClass = "astro-ph.CO",
    reportNumber = "IPPP/20/108",
    doi = "10.1088/1361-6382/ac086d",
    journal = "Class.\ Quant.\ Grav.",
    volume = "38",
    number = "15",
    pages = "153001",
    year = "2021"
}

@article{davis_hetdex_2023,
	title = {The {HETDEX} {Survey} {Emission}-line {Exploration} and {Source} {Classification}*},
	volume = {946},
	issn = {0004-637X, 1538-4357},
	url = {https://iopscience.iop.org/article/10.3847/1538-4357/acb0ca},
	doi = {10.3847/1538-4357/acb0ca},
	language = {en},
	number = {2},
	urldate = {2025-08-18},
	journal = {\apj},
	author = {Davis, Dustin and Gebhardt, Karl and Cooper, Erin Mentuch and Ciardullo, Robin and Fabricius, Maximilian and Farrow, Daniel J. and Feldmeier, John J. and Finkelstein, Steven L. and Gawiser, Eric and Gronwall, Caryl and Hill, Gary J. and Hopp, Ulrich and House, Lindsay R. and Jeong, Donghui and Kollatschny, Wolfram and Komatsu, Eiichiro and Landriau, Martin and Liu, Chenxu and Saito, Shun and Tuttle, Sarah and Wold, Isak G. B. and Zeimann, Gregory R. and Zhang, Yechi},
	month = apr,
	year = {2023},
	pages = {86},
}

@article{SDSS:2002vxn,
    author = "Blanton, Michael R. and others",
    collaboration = "SDSS",
    title = "{The Galaxy luminosity function and luminosity density at redshift z = 0.1}",
    eprint = "astro-ph/0210215",
    archivePrefix = "arXiv",
    doi = "10.1086/375776",
    journal = {\apj},
    volume = "592",
    pages = "819--838",
    year = "2003"
}

@misc{cross-parkin_dark_2025,
	title = {Dark sirens and the impact of redshift precision},
	url = {http://arxiv.org/abs/2502.17747},
	doi = {10.48550/arXiv.2502.17747},
	urldate = {2025-06-23},
	publisher = {arXiv},
	author = {Cross-Parkin, Madeline L. and Howlett, Cullan and Davis, Tamara M. and Khetan, Nandita},
	month = feb,
	year = {2025},
	note = {arXiv:2502.17747 [astro-ph]
version: 1},
}

@ARTICLE{het_catalog_1,
       author = {{Mentuch Cooper}, Erin and {Gebhardt}, Karl and {Davis}, Dustin and {Farrow}, Daniel J. and {Liu}, Chenxu and {Zeimann}, Gregory and {Ciardullo}, Robin and {Feldmeier}, John J. and {Drory}, Niv and {Jeong}, Donghui and {Benda}, Barbara and {Bowman}, William P. and {Boylan-Kolchin}, Michael and {Ch{\'a}vez Ortiz}, {\'O}scar A. and {Debski}, Maya H. and {Dentler}, Mona and {Fabricius}, Maximilian and {Farooq}, Rameen and {Finkelstein}, Steven L. and {Gawiser}, Eric and {Gronwall}, Caryl and {Hill}, Gary J. and {Hopp}, Ulrich and {House}, Lindsay R. and {Janowiecki}, Steven and {Khoraminezhad}, Hasti and {Kollatschny}, Wolfram and {Komatsu}, Eiichiro and {Landriau}, Martin and {Niemeyer}, Maja Lujan and {Lee}, Hanshin and {MacQueen}, Phillip and {Mawatari}, Ken and {McKay}, Brianna and {Ouchi}, Masami and {Poppe}, Jennifer and {Saito}, Shun and {Schneider}, Donald P. and {Snigula}, Jan and {Thomas}, Benjamin P. and {Tuttle}, Sarah and {Urrutia}, Tanya and {Weiss}, Laurel and {Wisotzki}, Lutz and {Zhang}, Yechi and {HETDEX Collaboration}},
        title = "{HETDEX Public Source Catalog 1: 220 K Sources Including Over 50 K Ly{\ensuremath{\alpha}} Emitters from an Untargeted Wide-area Spectroscopic Survey}",
      journal = {\apj},
         year = 2023,
        month = feb,
       volume = {943},
       number = {2},
          eid = {177},
        pages = {177},
          doi = {10.3847/1538-4357/aca962},
archivePrefix = {arXiv},
       eprint = {2301.01826},
 primaryClass = {astro-ph.GA},
       adsurl = {https://ui.adsabs.harvard.edu/abs/2023ApJ...943..177M}
}

@article{Harris_2020,
   title={Array programming with NumPy},
   volume={585},
   ISSN={1476-4687},
   url={http://dx.doi.org/10.1038/s41586-020-2649-2},
   DOI={10.1038/s41586-020-2649-2},
   number={7825},
   journal={Nature},
   publisher={Springer Science and Business Media LLC},
   author={Harris, Charles R. and Millman, K. Jarrod and van der Walt, Stéfan J. and Gommers, Ralf and Virtanen, Pauli and Cournapeau, David and Wieser, Eric and Taylor, Julian and Berg, Sebastian and Smith, Nathaniel J. and Kern, Robert and Picus, Matti and Hoyer, Stephan and van Kerkwijk, Marten H. and Brett, Matthew and Haldane, Allan and del Río, Jaime Fernández and Wiebe, Mark and Peterson, Pearu and Gérard-Marchant, Pierre and Sheppard, Kevin and Reddy, Tyler and Weckesser, Warren and Abbasi, Hameer and Gohlke, Christoph and Oliphant, Travis E.},
   year={2020},
   month=sep, pages={357–362} }

@article{Mastrogiovanni:2023emh,
    author = "Mastrogiovanni, Simone and Laghi, Danny and Gray, Rachel and Santoro, Giada Caneva and Ghosh, Archisman and Karathanasis, Christos and Leyde, Konstantin and Steer, Daniele A. and Perries, Stephane and Pierra, Gregoire",
    title = "{Joint population and cosmological properties inference with gravitational waves standard sirens and galaxy surveys}",
    eprint = "2305.10488",
    archivePrefix = "arXiv",
    primaryClass = "astro-ph.CO",
    doi = "10.1103/PhysRevD.108.042002",
    journal = "Phys. Rev. D",
    volume = "108",
    number = "4",
    pages = "042002",
    year = "2023"
}

@article{Chatterjee:2021xrm,
    author = "Chatterjee, Deep and Hegade K R, Abhishek and Holder, Gilbert and Holz, Daniel E. and Perkins, Scott and Yagi, Kent and Yunes, Nicol{\'a}s",
    title = "{Cosmology with Love: Measuring the Hubble constant using neutron star universal relations}",
    eprint = "2106.06589",
    archivePrefix = "arXiv",
    primaryClass = "gr-qc",
    doi = "10.1103/PhysRevD.104.083528",
    journal = "Phys. Rev. D",
    volume = "104",
    number = "8",
    pages = "083528",
    year = "2021"
}

@article{VanWyngarden:2025ogy,
    author = "VanWyngarden, Madison and Fishbach, Maya and Vijaykumar, Aditya and Guerrero, Alexandra G. and Holz, Daniel E.",
    title = "{How Low Can You Go: Constraining the Effects of Catalog Incompleteness on Dark Siren Cosmology}",
    eprint = "2511.04786",
    archivePrefix = "arXiv",
    primaryClass = "astro-ph.CO",
    month = "11",
    year = "2025"
}

@article{Pandey:2024mlo,
    author = "Pandey, Shiksha and Gupta, Ish and Chandra, Koustav and Sathyaprakash, Bangalore S.",
    title = "{The Critical Role of LIGO-India in the Era of Next-generation Observatories}",
    eprint = "2411.10349",
    archivePrefix = "arXiv",
    primaryClass = "gr-qc",
    doi = "10.3847/2041-8213/add15f",
    journal = "Astrophys. J. Lett.",
    volume = "985",
    number = "1",
    pages = "L17",
    year = "2025"
}

@misc{ligoindia,
    author = "Iyer, Bala and Souradeep, Tarun and Unnikrishnan, C. S. and Dhurandhar, Sanjeev and Raja, Sendhil and Kumar, Ajai and Sengupta, Anand",
    title = "{LIGO-India, Proposal of the Consortium for Indian Initiative in Gravitational-wave Observations (IndIGO)}",
    year = "2011",
    url = "https://dcc.ligo.org/LIGO-M1100296/public"
}

@article{KAGRA:2020tym,
    author = "Akutsu, T. and others",
    collaboration = "KAGRA",
    title = "{Overview of KAGRA: Detector design and construction history}",
    eprint = "2005.05574",
    archivePrefix = "arXiv",
    primaryClass = "physics.ins-det",
    doi = "10.1093/ptep/ptaa125",
    journal = "PTEP",
    volume = "2021",
    number = "5",
    pages = "05A101",
    year = "2021"
}

@article{KAGRA:2025dra,
    author = "Akutsu, T. and others",
    collaboration = "KAGRA",
    title = "{Decadal upgrade strategy for KAGRA toward post-O5 gravitational-wave astronomy}",
    eprint = "2508.03392",
    archivePrefix = "arXiv",
    primaryClass = "gr-qc",
    reportNumber = "JGW-P2516701",
    month = "8",
    year = "2025"
}

@article{Ghosh:2023ksl,
    author = "Ghosh, Tathagata and More, Surhud and Bera, Sayantani and Bose, Sukanta",
    title = "{Bayesian framework to infer the Hubble constant from the cross-correlation of individual gravitational wave events with galaxies}",
    eprint = "2312.16305",
    archivePrefix = "arXiv",
    primaryClass = "astro-ph.CO",
    reportNumber = "LIGO-P2300428",
    doi = "10.1103/PhysRevD.111.063513",
    journal = "Phys. Rev. D",
    volume = "111",
    number = "6",
    pages = "063513",
    year = "2025"
}

@inproceedings{Ghosh:2025qwc,
    author = "Ghosh, Tathagata and More, Surhud",
    title = "{Constraining the Hubble Constant using Cross-Correlation of Gravitational Wave Events with Flux-Limited Galaxy Catalog}",
    booktitle = "{24th International Conference on General Relativity and Gravitation (GR24) and 16th Edoardo Amaldi Conference on Gravitational (Amaldi16) Waves}",
    eprint = "2510.22187",
    archivePrefix = "arXiv",
    primaryClass = "astro-ph.CO",
    reportNumber = "JGW-P2516864",
    month = "10",
    year = "2025"
}

@article{4most,
  doi = {10.18727/0722-6691/5117},
  
  url = {https://doi.eso.org/10.18727/0722-6691/5117},
  
  author = {De Jong, Roelof S. and Agertz, Oscar and Berbel, Alex Agudo and Aird, James and Alexander, David A. and Amarsi, Anish and Anders, Friedrich and Andrae, Rene and Ansarinejad, Behzad and Ansorge, Wolfgang and Antilogus, Pierre and Anwand-Heerwart, Heiko and Arentsen, Anke and Arnadottir, Anna and Asplund, Martin and Auger, Matt and Azais, Nicolas and Baade, Dietrich and Baker, Gabriella and Baker, Sufyan and Balbinot, Eduardo and Baldry, Ivan K. and Banerji, Manda and Barden, Samuel and Barklem, Paul and Barthélémy-Mazot, Eléonore and Battistini, Chiara and Bauer, Svend and Bell, Cameron P. M. and Bellido-Tirado, Olga and Bellstedt, Sabine and Belokurov, Vasily and Bensby, Thomas and Bergemann, Maria and Bestenlehner, Joachim M. and Bielby, Richard and Bilicki, Maciej and Blake, Chris and Bland-Hawthorn, Joss and Boeche, Corrado and Boland, Wilfried and Boller, Thomas and Bongard, Sebastien and Bongiorno, Angela and Bonifacio, Piercarlo and Boudon, Didier and Brooks, David and Brown, Michael J. I. and Brown, Rebecca and Brüggen, Marcus and Brynnel, Joar and Brzeski, Jurek and Buchert, Thomas and Buschkamp, Peter and Caffau, Elisabetta and Caillier, Patrick and Carrick, Jonathan and Casagrande, Luca and Case, Scott and Casey, Andrew and Cesarini, Isabella and Cescutti, Gabriele and Chapuis, Diane and Chiappini, Cristina and Childress, Michael and Christlieb, Norbert and Church, Ross and Cioni, Maria-Rosa L. and Cluver, Michelle and Colless, Matthew and Collett, Thomas and Comparat, Johan and Cooper, Andrew and Couch, Warrick and Courbin, Frederic and Croom, Scott and Croton, Darren and Daguisé, Eric and Dalton, Gavin and Davies, Luke J. M. and Davis, Tamara and De Laverny, Patrick and Deason, Alis and Dionies, Frank and Disseau, Karen and Doel, Peter and Döscher, Daniel and Driver, Simon P. and Dwelly, Tom and Eckert, Dominique and Edge, Alastair and Edvardsson, Bengt and Youssoufi, Dalal El and Elhaddad, Ahmed and Enke, Harry and Erfanianfar, Ghazaleh and Farrell, Tony and Fechner, Thomas and Feiz, Carmen and Feltzing, Sofia and Ferreras, Ignacio and Feuerstein, Dietrich and Feuillet, Diane and Finoguenov, Alexis and Ford, Dominic and Fotopoulou, Sotiria and Fouesneau, Morgan and Frenk, Carlos and Frey, Steffen and Gaessler, Wolfgang and Geier, Stephan and Fusillo, Nicola Gentile and Gerhard, Ortwin and Giannantonio, Tommaso and Giannone, Domenico and Gibson, Brad and Gillingham, Peter and González-Fernández, Carlos and Gonzalez-Solares, Eduardo and Gottloeber, Stefan and Gould, Andrew and Grebel, Eva K. and Gueguen, Alain and Guiglion, Guillaume and Haehnelt, Martin and Hahn, Thomas and Hansen, Camilla J. and Hartman, Henrik and Hauptner, Katja and Hawkins, Keith and Haynes, Dionne and Haynes, Roger and Heiter, Ulrike and Helmi, Amina and Aguayo, Cesar Hernandez and Hewett, Paul and Hinton, Samuel and Hobbs, David and Hoenig, Sebastian and Hofman, David and Hook, Isobel and Hopgood, Joshua and Hopkins, Andrew and Hourihane, Anna and Howes, Louise and Howlett, Cullan and Huet, Tristan and Irwin, Mike and Iwert, Olaf and Jablonka, Pascale and Jahn, Thomas and Jahnke, Knud and Jarno, Aurélien and Jin, Shoko and Jofre, Paula and Johl, Diana and Jones, Damien and Jönsson, Henrik and Jordan, Carola and Karovicova, Iva and Khalatyan, Arman and Kelz, Andreas and Kennicutt, Robert and King, David and Kitaura, Francisco and Klar, Jochen and Klauser, Urs and Kneib, Jean-Paul and Koch, Andreas and Koposov, Sergey and Kordopatis, Georges and Korn, Andreas and Kosmalski, Johan and Kotak, Rubina and Kovalev, Mikhail and Kreckel, Kathryn and Kripak, Yevgen and Krumpe, Mirko and Kuijken, Koen and Kunder, Andrea and Kushniruk, Iryna and Lam, Man I and Lamer, Georg and Laurent, Florence and Lawrence, Jon and Lehmitz, Michael and Lemasle, Bertrand and Lewis, James and Li, Baojiu and Lidman, Chris and Lind, Karin and Liske, Jochen and Lizon, Jean-Louis and Loveday, Jon and Ludwig, Hans-Günter and McDermid, Richard M. and Maguire, Kate and Mainieri, Vincenzo and Mali, Slavko and Mandel, Holger and Mandel, Kaisey and Mannering, Liz and Martell, Sarah and Delgado, David Martinez and Matijevic, Gal and McGregor, Helen and McMahon, Richard and McMillan, Paul and Mena, Olga and Merloni, Andrea and Meyer, Martin J. and Michel, Christophe and Micheva, Genoveva and Migniau, Jean-Emmanuel and Minchev, Ivan and Monari, Giacomo and Muller, Rolf and Murphy, David and Muthukrishna, Daniel and Nandra, Kirpal and Navarro, Ramon and Ness, Melissa and Nichani, Vijay and Nichol, Robert and Nicklas, Harald and Niederhofer, Florian and Norberg, Peder and Obreschkow, Danail and Oliver, Seb and Owers, Matt and Pai, Naveen and Pankratow, Sergei and Parkinson, David and Paschke, Jens and Paterson, Robert and Pecontal, Arlette and Parry, Ian and Phillips, Dan and Pillepich, Annalisa and Pinard, Laurent and Pirard, Jeff and Piskunov, Nikolai and Plank, Volker and Plüschke, Dennis and Pons, Estelle and Popesso, Paola and Power, Chris and Pragt, Johan and Pramskiy, Alexander and Pryer, Dan and Quattri, Marco and Queiroz, Anna Barbara De Andrade and Quirrenbach, Andreas and Rahurkar, Swara and Raichoor, Anand and Ramstedt, Sofia and Rau, Arne and Recio-Blanco, Alejandra and Reiss, Roland and Renaud, Florent and Revaz, Yves and Rhode, Petra and Richard, Johan and Richter, Amon David and Rix, Hans-Walter and Robotham, Aaron S. G. and Roelfsema, Ronald and Romaniello, Martino and Rosario, David and Rothmaier, Florian and Roukema, Boudewijn and Ruchti, Gregory and Rupprecht, Gero and Rybizki, Jan and Ryde, Nils and Saar, Andre and Sadler, Elaine and Sahlén, Martin and Salvato, Mara and Sassolas, Benoit and Saunders, Will and Saviauk, Allar and Sbordone, Luca and Schmidt, Thomas and Schnurr, Olivier and Scholz, Ralf-Dieter and Schwope, Axel and Seifert, Walter and Shanks, Tom and Sheinis, Andrew and Sivov, Tihomir and Skúladóttir, Ása and Smartt, Stephen and Smedley, Scott and Smith, Greg and Smith, Robert and Sorce, Jenny and Spitler, Lee and Starkenburg, Else and Steinmetz, Matthias and Stilz, Ingo and Storm, Jesper and Sullivan, Mark and Sutherland, William and Swann, Elizabeth and Tamone, Amélie and Taylor, Edward N. and Teillon, Julien and Tempel, Elmo and Ter Horst, Rik and Thi, Wing-Fai and Tolstoy, Eline and Trager, Scott and Traven, Gregor and Tremblay, Pier-Emmanuel and Tresse, Laurence and Valentini, Marica and Van De Weygaert, Rien and Van Den Ancker, Mario and Veljanoski, Jovan and Venkatesan, Sudharshan and Wagner, Lukas and Wagner, Karl and Walcher, C. Jakob and Waller, Lew and Walton, Nicholas and Wang, Lingyu and Winkler, Roland and Wisotzki, Lutz and Worley, C. Clare and Worseck, Gabor and Xiang, Maosheng and Xu, Wenli and Yong, David and Zhao, Cheng and Zheng, Jessica and Zscheyge, Florian and Zucker, Daniel},
  
  title = {4MOST: Project overview and information for the First Call for Proposals},
  
  journal = {Published in The Messenger vol. 175},
  
  volume = {pp. 3-11},
  
  pages = {March 2019.},
  
  publisher = {European Southern Observatory (ESO)},
  
  year = {2019},
  
  copyright = {Copyright European Southern Observatory}
}

@article{LIGOScientific:2025slb,
    author = "Abac, A. G. and others",
    collaboration = "LIGO Scientific, VIRGO, KAGRA",
    title = "{GWTC-4.0: Updating the Gravitational-Wave Transient Catalog with Observations from the First Part of the Fourth LIGO-Virgo-KAGRA Observing Run}",
    eprint = "2508.18082",
    archivePrefix = "arXiv",
    primaryClass = "gr-qc",
    reportNumber = "LIGO-P2400386",
    month = "8",
    year = "2025"
}

@article{Kalomenopoulos:2025qpt,
    author = "Kalomenopoulos, Marios and Barbieri, Riccardo and Khochfar, Sadegh and Gair, Jonathan and McGibbon, Robert J.",
    title = "{Clustering effects on the Dark Siren determination of $H_0$: A simulation study}",
    eprint = "2511.12334",
    archivePrefix = "arXiv",
    primaryClass = "astro-ph.CO",
    month = "11",
    year = "2025"
}

@article{Messenger:2011gi,
    author = "Messenger, C. and Read, J.",
    title = "{Measuring a cosmological distance-redshift relationship using only gravitational wave observations of binary neutron star coalescences}",
    eprint = "1107.5725",
    archivePrefix = "arXiv",
    primaryClass = "gr-qc",
    doi = "10.1103/PhysRevLett.108.091101",
    journal = "Phys. Rev. Lett.",
    volume = "108",
    pages = "091101",
    year = "2012"
}

@article{Dhani:2022ulg,
    author = "Dhani, Arnab and Borhanian, Ssohrab and Gupta, Anuradha and Sathyaprakash, Bangalore",
    title = "{Cosmography with bright and Love sirens}",
    eprint = "2212.13183",
    archivePrefix = "arXiv",
    primaryClass = "gr-qc",
    month = "12",
    year = "2022"
}

@article{Zhan:2025jqg,
    author = "Zhan, Yejing and Izquierdo-Villalba, David and Guo, Xiao and Yang, Qing and Spinoso, Daniele and Wang, Fa-Yin",
    title = "{Bright Siren without an Electromagnetic Counterpart Detected by the LISA-Taiji-TianQin Network}",
    eprint = "2509.04218",
    archivePrefix = "arXiv",
    primaryClass = "astro-ph.CO",
    doi = "10.3847/1538-4357/ae1743",
    journal = "Astrophys. J.",
    volume = "995",
    number = "1",
    pages = "71",
    year = "2025"
}

@article{Driver:2022vyh,
    author = "Driver, Simon P. and others",
    title = "{Galaxy And Mass Assembly (GAMA): Data Release 4 and the z {\ensuremath{<}} 0.1 total and z {\ensuremath{<}} 0.08 morphological galaxy stellar mass functions}",
    eprint = "2203.08539",
    archivePrefix = "arXiv",
    primaryClass = "astro-ph.GA",
    doi = "10.1093/mnras/stac472",
    journal = "Mon. Not. Roy. Astron. Soc.",
    volume = "513",
    number = "1",
    pages = "439--467",
    year = "2022"
}

@article{Mukherjee:2020hyn,
    author = "Mukherjee, Suvodip and Wandelt, Benjamin D. and Nissanke, Samaya M. and Silvestri, Alessandra",
    title = "{Accurate precision Cosmology with redshift unknown gravitational wave sources}",
    eprint = "2007.02943",
    archivePrefix = "arXiv",
    primaryClass = "astro-ph.CO",
    doi = "10.1103/PhysRevD.103.043520",
    journal = "Phys. Rev. D",
    volume = "103",
    number = "4",
    pages = "043520",
    year = "2021"
}

@article{Afroz:2024joi,
    author = "Afroz, Samsuzzaman and Mukherjee, Suvodip",
    title = "{Prospect of precision cosmology and testing general relativity using binary black holes {\textendash} galaxies cross-correlation}",
    eprint = "2407.09262",
    archivePrefix = "arXiv",
    primaryClass = "astro-ph.CO",
    doi = "10.1093/mnras/stae2139",
    journal = "Mon. Not. Roy. Astron. Soc.",
    volume = "534",
    number = "2",
    pages = "1283--1298",
    year = "2024"
}

@article{Mukherjee:2019wcg,
    author = "Mukherjee, Suvodip and Wandelt, Benjamin D. and Silk, Joseph",
    title = "{Probing the theory of gravity with gravitational lensing of gravitational waves and galaxy surveys}",
    eprint = "1908.08951",
    archivePrefix = "arXiv",
    primaryClass = "astro-ph.CO",
    doi = "10.1093/mnras/staa827",
    journal = "Mon. Not. Roy. Astron. Soc.",
    volume = "494",
    number = "2",
    pages = "1956--1970",
    year = "2020"
}

@ARTICLE{2007matplotlib,
       author = {{Hunter}, John D.},
        title = "{Matplotlib: A 2D Graphics Environment}",
      journal = {Computing in Science and Engineering},
         year = 2007,
        month = jan,
       volume = {9},
       number = {3},
        pages = {90-95},
          doi = {10.1109/MCSE.2007.55},
       adsurl = {https://ui.adsabs.harvard.edu/abs/2007CSE.....9...90H}
}

@article{astropy2018,
   title={The Astropy Project: Building an Open-science Project and Status of the v2.0 Core Package*},
   volume={156},
   ISSN={1538-3881},
   url={http://dx.doi.org/10.3847/1538-3881/aabc4f},
   DOI={10.3847/1538-3881/aabc4f},
   number={3},
   journal={The Astronomical Journal},
   publisher={American Astronomical Society},
   author={Price-Whelan, A. M. and Sipőcz, B. M. and Günther, H. M. and Lim, P. L. and Crawford, S. M. and Conseil, S. and Shupe, D. L. and Craig, M. W. and Dencheva, N. and Ginsburg, A. and VanderPlas, J. T. and Bradley, L. D. and Pérez-Suárez, D. and de Val-Borro, M. and Aldcroft, T. L. and Cruz, K. L. and Robitaille, T. P. and Tollerud, E. J. and Ardelean, C. and Babej, T. and Bach, Y. P. and Bachetti, M. and Bakanov, A. V. and Bamford, S. P. and Barentsen, G. and Barmby, P. and Baumbach, A. and Berry, K. L. and Biscani, F. and Boquien, M. and Bostroem, K. A. and Bouma, L. G. and Brammer, G. B. and Bray, E. M. and Breytenbach, H. and Buddelmeijer, H. and Burke, D. J. and Calderone, G. and Rodríguez, J. L. Cano and Cara, M. and Cardoso, J. V. M. and Cheedella, S. and Copin, Y. and Corrales, L. and Crichton, D. and D’Avella, D. and Deil, C. and Depagne, É. and Dietrich, J. P. and Donath, A. and Droettboom, M. and Earl, N. and Erben, T. and Fabbro, S. and Ferreira, L. A. and Finethy, T. and Fox, R. T. and Garrison, L. H. and Gibbons, S. L. J. and Goldstein, D. A. and Gommers, R. and Greco, J. P. and Greenfield, P. and Groener, A. M. and Grollier, F. and Hagen, A. and Hirst, P. and Homeier, D. and Horton, A. J. and Hosseinzadeh, G. and Hu, L. and Hunkeler, J. S. and Ivezić, Ž. and Jain, A. and Jenness, T. and Kanarek, G. and Kendrew, S. and Kern, N. S. and Kerzendorf, W. E. and Khvalko, A. and King, J. and Kirkby, D. and Kulkarni, A. M. and Kumar, A. and Lee, A. and Lenz, D. and Littlefair, S. P. and Ma, Z. and Macleod, D. M. and Mastropietro, M. and McCully, C. and Montagnac, S. and Morris, B. M. and Mueller, M. and Mumford, S. J. and Muna, D. and Murphy, N. A. and Nelson, S. and Nguyen, G. H. and Ninan, J. P. and Nöthe, M. and Ogaz, S. and Oh, S. and Parejko, J. K. and Parley, N. and Pascual, S. and Patil, R. and Patil, A. A. and Plunkett, A. L. and Prochaska, J. X. and Rastogi, T. and Janga, V. Reddy and Sabater, J. and Sakurikar, P. and Seifert, M. and Sherbert, L. E. and Sherwood-Taylor, H. and Shih, A. Y. and Sick, J. and Silbiger, M. T. and Singanamalla, S. and Singer, L. P. and Sladen, P. H. and Sooley, K. A. and Sornarajah, S. and Streicher, O. and Teuben, P. and Thomas, S. W. and Tremblay, G. R. and Turner, J. E. H. and Terrón, V. and Kerkwijk, M. H. van and de la Vega, A. and Watkins, L. L. and Weaver, B. A. and Whitmore, J. B. and Woillez, J. and Zabalza, V.},
   year={2018},
   month=aug, pages={123} }

@article{Virtanen_2020,
   title={SciPy 1.0: fundamental algorithms for scientific computing in Python},
   volume={17},
   ISSN={1548-7105},
   url={http://dx.doi.org/10.1038/s41592-019-0686-2},
   DOI={10.1038/s41592-019-0686-2},
   number={3},
   journal={Nature Methods},
   publisher={Springer Science and Business Media LLC},
   author={Virtanen, Pauli and Gommers, Ralf and Oliphant, Travis E. and Haberland, Matt and Reddy, Tyler and Cournapeau, David and Burovski, Evgeni and Peterson, Pearu and Weckesser, Warren and Bright, Jonathan and van der Walt, Stéfan J. and Brett, Matthew and Wilson, Joshua and Millman, K. Jarrod and Mayorov, Nikolay and Nelson, Andrew R. J. and Jones, Eric and Kern, Robert and Larson, Eric and Carey, C J and Polat, İlhan and Feng, Yu and Moore, Eric W. and VanderPlas, Jake and Laxalde, Denis and Perktold, Josef and Cimrman, Robert and Henriksen, Ian and Quintero, E. A. and Harris, Charles R. and Archibald, Anne M. and Ribeiro, Antônio H. and Pedregosa, Fabian and van Mulbregt, Paul and Vijaykumar, Aditya and Bardelli, Alessandro Pietro and Rothberg, Alex and Hilboll, Andreas and Kloeckner, Andreas and Scopatz, Anthony and Lee, Antony and Rokem, Ariel and Woods, C. Nathan and Fulton, Chad and Masson, Charles and Häggström, Christian and Fitzgerald, Clark and Nicholson, David A. and Hagen, David R. and Pasechnik, Dmitrii V. and Olivetti, Emanuele and Martin, Eric and Wieser, Eric and Silva, Fabrice and Lenders, Felix and Wilhelm, Florian and Young, G. and Price, Gavin A. and Ingold, Gert-Ludwig and Allen, Gregory E. and Lee, Gregory R. and Audren, Hervé and Probst, Irvin and Dietrich, Jörg P. and Silterra, Jacob and Webber, James T and Slavič, Janko and Nothman, Joel and Buchner, Johannes and Kulick, Johannes and Schönberger, Johannes L. and de Miranda Cardoso, José Vinícius and Reimer, Joscha and Harrington, Joseph and Rodríguez, Juan Luis Cano and Nunez-Iglesias, Juan and Kuczynski, Justin and Tritz, Kevin and Thoma, Martin and Newville, Matthew and Kümmerer, Matthias and Bolingbroke, Maximilian and Tartre, Michael and Pak, Mikhail and Smith, Nathaniel J. and Nowaczyk, Nikolai and Shebanov, Nikolay and Pavlyk, Oleksandr and Brodtkorb, Per A. and Lee, Perry and McGibbon, Robert T. and Feldbauer, Roman and Lewis, Sam and Tygier, Sam and Sievert, Scott and Vigna, Sebastiano and Peterson, Stefan and More, Surhud and Pudlik, Tadeusz and Oshima, Takuya and Pingel, Thomas J. and Robitaille, Thomas P. and Spura, Thomas and Jones, Thouis R. and Cera, Tim and Leslie, Tim and Zito, Tiziano and Krauss, Tom and Upadhyay, Utkarsh and Halchenko, Yaroslav O. and Vázquez-Baeza, Yoshiki},
   year={2020},
   month=feb, pages={261–272} }

@ARTICLE{het_descrip,
       author = {{Gebhardt}, Karl and {Mentuch Cooper}, Erin and {Ciardullo}, Robin and {Acquaviva}, Viviana and {Bender}, Ralf and {Bowman}, William P. and {Castanheira}, Barbara G. and {Dalton}, Gavin and {Davis}, Dustin and {de Jong}, Roelof S. and {DePoy}, D.~L. and {Devarakonda}, Yaswant and {Dongsheng}, Sun and {Drory}, Niv and {Fabricius}, Maximilian and {Farrow}, Daniel J. and {Feldmeier}, John and {Finkelstein}, Steven L. and {Froning}, Cynthia S. and {Gawiser}, Eric and {Gronwall}, Caryl and {Herold}, Laura and {Hill}, Gary J. and {Hopp}, Ulrich and {House}, Lindsay R. and {Janowiecki}, Steven and {Jarvis}, Matthew and {Jeong}, Donghui and {Jogee}, Shardha and {Kakuma}, Ryota and {Kelz}, Andreas and {Kollatschny}, W. and {Komatsu}, Eiichiro and {Krumpe}, Mirko and {Landriau}, Martin and {Liu}, Chenxu and {Niemeyer}, Maja Lujan and {MacQueen}, Phillip and {Marshall}, Jennifer and {Mawatari}, Ken and {McLinden}, Emily M. and {Mukae}, Shiro and {Nagaraj}, Gautam and {Ono}, Yoshiaki and {Ouchi}, Masami and {Papovich}, Casey and {Sakai}, Nao and {Saito}, Shun and {Schneider}, Donald P. and {Schulze}, Andreas and {Shanmugasundararaj}, Khavvia and {Shetrone}, Matthew and {Sneden}, Chris and {Snigula}, Jan and {Steinmetz}, Matthias and {Thomas}, Benjamin P. and {Thomas}, Brianna and {Tuttle}, Sarah and {Urrutia}, Tanya and {Wisotzki}, Lutz and {Wold}, Isak and {Zeimann}, Gregory and {Zhang}, Yechi},
        title = "{The Hobby-Eberly Telescope Dark Energy Experiment (HETDEX) Survey Design, Reductions, and Detections}",
      journal = {\apj},
         year = 2021,
        month = dec,
       volume = {923},
       number = {2},
          eid = {217},
        pages = {217},
          doi = {10.3847/1538-4357/ac2e03},
archivePrefix = {arXiv},
       eprint = {2110.04298},
 primaryClass = {astro-ph.IM},
       adsurl = {https://ui.adsabs.harvard.edu/abs/2021ApJ...923..217G}
}

@article{Schutz:1986gp,
    author = "Schutz, Bernard F.",
    title = "{Determining the Hubble Constant from Gravitational Wave Observations}",
    doi = "10.1038/323310a0",
    journal = {\nat},
    volume = "323",
    pages = "310--311",
    year = "1986"
}

@article{Holz:2005df,
    author = "Holz, Daniel E. and Hughes, Scott A.",
    title = "{Using gravitational-wave standard sirens}",
    eprint = "astro-ph/0504616",
    archivePrefix = "arXiv",
    doi = "10.1086/431341",
    journal = {\apj},
    volume = "629",
    pages = "15--22",
    year = "2005"
}

@unpublished{DarkEnergySurveyGravitationalWave:2025ykv,
    author={Haibin Zhang and Mitsuru Kokubo and Sean MacBride and Isaac McMahon and Nozomu Tominaga and Yousuke Utsumi and Michitoshi Yoshida and Tomoki Morokuma and Masaomi Tanaka and Akira Arai and Wanqiu He and Yuki Moritani and Masato Onodera and Vera Maria Passegger and Ichi Tanaka and Kiyoto Yabe and Lillian Joseph and Simran Kaur and Hemanth Bommireddy and Nora Sherman and Kenneth Herner and H. Thomas Diehl and Marcelle Soares-Santos},
    collaboration = "Dark Energy Survey Gravitational Wave, J-GEM",
    title = "{A Joint Search for the Electromagnetic Counterpart to the Gravitational-Wave Binary Black-Hole Merger Candidate S250328ae with the Dark Energy Camera and the Prime Focus Spectrograph}",
    eprint = "2508.00291",
    archivePrefix = "arXiv",
    primaryClass = "astro-ph.HE",
    reportNumber = "FERMILAB-PUB-25-0541-CSAID-PPD",
    month = "8",
    year = "2025",
    note = {\url{https://arxiv.org/pdf/2508.00291}}
}

@ARTICLE{Chen:2025qsl,
       author = {{Chen}, Anson},
        title = "{Measuring the cosmic dipole with golden dark sirens in the era of next-generation ground-based gravitational wave detectors}",
      journal = {\jcap},
         year = 2025,
        month = jul,
       volume = {2025},
       number = {7},
          eid = {076},
        pages = {076},
          doi = {10.1088/1475-7516/2025/07/076},
archivePrefix = {arXiv},
       eprint = {2505.12678},
 primaryClass = {gr-qc},
       adsurl = {https://ui.adsabs.harvard.edu/abs/2025JCAP...07..076C}
}

@article{Gray:2019ksv,
    author = "Gray, Rachel and others",
    title = "{Cosmological inference using gravitational wave standard sirens: A mock data analysis}",
    eprint = "1908.06050",
    archivePrefix = "arXiv",
    primaryClass = "gr-qc",
    reportNumber = "LIGO-P1900017",
    doi = "10.1103/PhysRevD.101.122001",
    journal = {\prd},
    volume = "101",
    number = "12",
    pages = "122001",
    year = "2020"
}

@article{gair_hitchhikers_2023,
	title = {The {Hitchhiker}'s guide to the galaxy catalog approach for gravitational wave cosmology},
	volume = {166},
	issn = {0004-6256, 1538-3881},
	url = {http://arxiv.org/abs/2212.08694},
	doi = {10.3847/1538-3881/acca78},
	number = {1},
	urldate = {2025-04-02},
	journal = {\aj},
	author = {Gair, Jonathan R. and Ghosh, Archisman and Gray, Rachel and Holz, Daniel E. and Mastrogiovanni, Simone and Mukherjee, Suvodip and Palmese, Antonella and Tamanini, Nicola and Baker, Tessa and Beirnaert, Freija and Bilicki, Maciej and Chen, Hsin-Yu and Dálya, Gergely and Ezquiaga, Jose Maria and Farr, Will M. and Fishbach, Maya and Garcia-Bellido, Juan and Ghosh, Tathagata and Huang, Hsiang-Yu and Karathanasis, Christos and Leyde, Konstantin and Hernandez, Ignacio Magaña and Noller, Johannes and Pierra, Gregoire and Raffai, Peter and Romano, Antonio Enea and Seglar-Arroyo, Monica and Steer, Danièle A. and Turski, Cezary and Vaccaro, Maria Paola and Vallejo-Peña, Sergio Andrés},
	month = jul,
	year = {2023},
	note = {arXiv:2212.08694 [gr-qc]},
	pages = {22},
}

@ARTICLE{Planck:2018vyg,
    author = "Aghanim, N. and others",
    collaboration = "Planck",
    title = "{Planck 2018 results. VI. Cosmological parameters}",
    eprint = "1807.06209",
    archivePrefix = "arXiv",
    primaryClass = "astro-ph.CO",
    doi = "10.1051/0004-6361/201833910",
    journal = {\aap},
    volume = "641",
    pages = "A6",
    year = "2020",
    note = "[Erratum: Astron.Astrophys. 652, C4 (2021)]"
}

@ARTICLE{abbott_gravitational-wave_2017,
	title = {A gravitational-wave standard siren measurement of the {Hubble} constant},
	volume = {551},
	issn = {1476-4687},
	url = {https://doi.org/10.1038/nature24471},
	doi = {10.1038/nature24471},
	number = {7678},
	journal = {\nat},
	author = {Abbott, B. P. and Abbott, R. and Abbott, T. D. and Acernese, F. and Ackley, K. and Adams, C. and Adams, T. and Addesso, P. and Adhikari, R. X. and Adya, V. B. and Affeldt, C. and Afrough, M. and Agarwal, B. and Agathos, M. and Agatsuma, K. and Aggarwal, N. and Aguiar, O. D. and Aiello, L. and Ain, A. and Ajith, P. and Allen, B. and Allen, G. and Allocca, A. and Altin, P. A. and Amato, A. and Ananyeva, A. and Anderson, S. B. and Anderson, W. G. and Angelova, S. V. and Antier, S. and Appert, S. and Arai, K. and Araya, M. C. and Areeda, J. S. and Arnaud, N. and Arun, K. G. and Ascenzi, S. and Ashton, G. and Ast, M. and Aston, S. M. and Astone, P. and Atallah, D. V. and Aufmuth, P. and Aulbert, C. and AultONeal, K. and Austin, C. and Avila-Alvarez, A. and Babak, S. and Bacon, P. and Bader, M. K. M. and Bae, S. and Baker, P. T. and Baldaccini, F. and Ballardin, G. and Ballmer, S. W. and Banagiri, S. and Barayoga, J. C. and Barclay, S. E. and Barish, B. C. and Barker, D. and Barkett, K. and Barone, F. and Barr, B. and Barsotti, L. and Barsuglia, M. and Barta, D. and Bartlett, J. and Bartos, I. and Bassiri, R. and Basti, A. and Batch, J. C. and Bawaj, M. and Bayley, J. C. and Bazzan, M. and Bécsy, B. and Beer, C. and Bejger, M. and Belahcene, I. and Bell, A. S. and Berger, B. K. and Bergmann, G. and Bero, J. J. and Berry, C. P. L. and Bersanetti, D. and Bertolini, A. and Betzwieser, J. and Bhagwat, S. and Bhandare, R. and Bilenko, I. A. and Billingsley, G. and Billman, C. R. and Birch, J. and Birney, R. and Birnholtz, O. and Biscans, S. and Biscoveanu, S. and Bisht, A. and Bitossi, M. and Biwer, C. and Bizouard, M. A. and Blackburn, J. K. and Blackman, J. and Blair, C. D. and Blair, D. G. and Blair, R. M. and Bloemen, S. and Bock, O. and Bode, N. and Boer, M. and Bogaert, G. and Bohe, A. and Bondu, F. and Bonilla, E. and Bonnand, R. and Boom, B. A. and Bork, R. and Boschi, V. and Bose, S. and Bossie, K. and Bouffanais, Y. and Bozzi, A. and Bradaschia, C. and Brady, P. R. and Branchesi, M. and Brau, J. E. and Briant, T. and Brillet, A. and Brinkmann, M. and Brisson, V. and Brockill, P. and Broida, J. E. and Brooks, A. F. and Brown, D. A. and Brown, D. D. and Brunett, S. and Buchanan, C. C. and Buikema, A. and Bulik, T. and Bulten, H. J. and Buonanno, A. and Buskulic, D. and Buy, C. and Byer, R. L. and Cabero, M. and Cadonati, L. and Cagnoli, G. and Cahillane, C. and Bustillo, J. Calderón and Callister, T. A. and Calloni, E. and Camp, J. B. and Canepa, M. and Canizares, P. and Cannon, K. C. and Cao, H. and Cao, J. and Capano, C. D. and Capocasa, E. and Carbognani, F. and Caride, S. and Carney, M. F. and Diaz, J. Casanueva and Casentini, C. and Caudill, S. and Cavaglià, M. and Cavalier, F. and Cavalieri, R. and Cella, G. and Cepeda, C. B. and Cerdá-Durán, P. and Cerretani, G. and Cesarini, E. and Chamberlin, S. J. and Chan, M. and Chao, S. and Charlton, P. and Chase, E. and Chassande-Mottin, E. and Chatterjee, D. and Chatziioannou, K. and Cheeseboro, B. D. and Chen, H. Y. and Chen, X. and Chen, Y. and Cheng, H.-P. and Chia, H. and Chincarini, A. and Chiummo, A. and Chmiel, T. and Cho, H. S. and Cho, M. and Chow, J. H. and Christensen, N. and Chu, Q. and Chua, A. J. K. and Chua, S. and Chung, A. K. W. and Chung, S. and Ciani, G. and Ciolfi, R. and Cirelli, C. E. and Cirone, A. and Clara, F. and Clark, J. A. and Clearwater, P. and Cleva, F. and Cocchieri, C. and Coccia, E. and Cohadon, P.-F. and Cohen, D. and Colla, A. and Collette, C. G. and Cominsky, L. R. and Constancio, M. and Conti, L. and Cooper, S. J. and Corban, P. and Corbitt, T. R. and Cordero-Carrión, I. and Corley, K. R. and Cornish, N. and Corsi, A. and Cortese, S. and Costa, C. A. and Coughlin, M. W. and Coughlin, S. B. and Coulon, J.-P. and Countryman, S. T. and Couvares, P. and Covas, P. B. and Cowan, E. E. and Coward, D. M. and Cowart, M. J. and Coyne, D. C. and Coyne, R. and Creighton, J. D. E. and Creighton, T. D. and Cripe, J. and Crowder, S. G. and Cullen, T. J. and Cumming, A. and Cunningham, L. and Cuoco, E. and Dal Canton, T. and Dálya, G. and Danilishin, S. L. and D’Antonio, S. and Danzmann, K. and Dasgupta, A. and Da Silva Costa, C. F. and Datrier, L. E. H. and Dattilo, V. and Dave, I. and Davier, M. and Davis, D. and Daw, E. J. and Day, B. and De, S. and DeBra, D. and Degallaix, J. and De Laurentis, M. and Deléglise, S. and Del Pozzo, W. and Demos, N. and Denker, T. and Dent, T. and De Pietri, R. and Dergachev, V. and De Rosa, R. and DeRosa, R. T. and De Rossi, C. and DeSalvo, R. and de Varona, O. and Devenson, J. and Dhurandhar, S. and Díaz, M. C. and Di Fiore, L. and Di Giovanni, M. and Di Girolamo, T. and Di Lieto, A. and Di Pace, S. and Di Palma, I. and Di Renzo, F. and Doctor, Z. and Dolique, V. and Donovan, F. and Dooley, K. L. and Doravari, S. and Dorrington, I. and Douglas, R. and Dovale álvarez, M. and Downes, T. P. and Drago, M. and Dreissigacker, C. and Driggers, J. C. and Du, Z. and Ducrot, M. and Dupej, P. and Dwyer, S. E. and {The LIGO Scientific Collaboration and The Virgo Collaboration}},
	month = nov,
	year = {2017},
	pages = {85--88},
}

@ARTICLE{collaboration_advanced_2015,
	title = {Advanced {LIGO}},
	volume = {32},
	url = {https://dx.doi.org/10.1088/0264-9381/32/7/074001},
	doi = {10.1088/0264-9381/32/7/074001},
	number = {7},
	journal = {Class.\ Quant.\ Gravity},
	author = {Collaboration, The LIGO Scientific and Aasi, J and Abbott, B P and Abbott, R and Abbott, T and Abernathy, M R and Ackley, K and Adams, C and Adams, T and Addesso, P and Adhikari, R X and Adya, V and Affeldt, C and Aggarwal, N and Aguiar, O D and Ain, A and Ajith, P and Alemic, A and Allen, B and Amariutei, D and Anderson, S B and Anderson, W G and Arai, K and Araya, M C and Arceneaux, C and Areeda, J S and Ashton, G and Ast, S and Aston, S M and Aufmuth, P and Aulbert, C and Aylott, B E and Babak, S and Baker, P T and Ballmer, S W and Barayoga, J C and Barbet, M and Barclay, S and Barish, B C and Barker, D and Barr, B and Barsotti, L and Bartlett, J and Barton, M A and Bartos, I and Bassiri, R and Batch, J C and Baune, C and Behnke, B and Bell, A S and Bell, C and Benacquista, M and Bergman, J and Bergmann, G and Berry, C P L and Betzwieser, J and Bhagwat, S and Bhandare, R and Bilenko, I A and Billingsley, G and Birch, J and Biscans, S and Biwer, C and Blackburn, J K and Blackburn, L and Blair, C D and Blair, D and Bock, O and Bodiya, T P and Bojtos, P and Bond, C and Bork, R and Born, M and Bose, Sukanta and Brady, P R and Braginsky, V B and Brau, J E and Bridges, D O and Brinkmann, M and Brooks, A F and Brown, D A and Brown, D D and Brown, N M and Buchman, S and Buikema, A and Buonanno, A and Cadonati, L and Calderón Bustillo, J and Camp, J B and Cannon, K C and Cao, J and Capano, C D and Caride, S and Caudill, S and Cavaglià, M and Cepeda, C and Chakraborty, R and Chalermsongsak, T and Chamberlin, S J and Chao, S and Charlton, P and Chen, Y and Cho, H S and Cho, M and Chow, J H and Christensen, N and Chu, Q and Chung, S and Ciani, G and Clara, F and Clark, J A and Collette, C and Cominsky, L and Constancio, M and Cook, D and Corbitt, T R and Cornish, N and Corsi, A and Costa, C A and Coughlin, M W and Countryman, S and Couvares, P and Coward, D M and Cowart, M J and Coyne, D C and Coyne, R and Craig, K and Creighton, J D E and Creighton, T D and Cripe, J and Crowder, S G and Cumming, A and Cunningham, L and Cutler, C and Dahl, K and Dal Canton, T and Damjanic, M and Danilishin, S L and Danzmann, K and Dartez, L and Dave, I and Daveloza, H and Davies, G S and Daw, E J and DeBra, D and Del Pozzo, W and Denker, T and Dent, T and Dergachev, V and DeRosa, R T and DeSalvo, R and Dhurandhar, S and D´ıaz, M and Di Palma, I and Dojcinoski, G and Dominguez, E and Donovan, F and Dooley, K L and Doravari, S and Douglas, R and Downes, T P and Driggers, J C and Du, Z and Dwyer, S and Eberle, T and Edo, T and Edwards, M and Edwards, M and Effler, A and Eggenstein, H.-B and Ehrens, P and Eichholz, J and Eikenberry, S S and Essick, R and Etzel, T and Evans, M and Evans, T and Factourovich, M and Fairhurst, S and Fan, X and Fang, Q and Farr, B and Farr, W M and Favata, M and Fays, M and Fehrmann, H and Fejer, M M and Feldbaum, D and Ferreira, E C and Fisher, R P and Frei, Z and Freise, A and Frey, R and Fricke, T T and Fritschel, P and Frolov, V V and Fuentes-Tapia, S and Fulda, P and Fyffe, M and Gair, J R and Gaonkar, S and Gehrels, N and Gergely´, L Á and Giaime, J A and Giardina, K D and Gleason, J and Goetz, E and Goetz, R and Gondan, L and González, G and Gordon, N and Gorodetsky, M L and Gossan, S and Goßler, S and Gräf, C and Graff, P B and Grant, A and Gras, S and Gray, C and Greenhalgh, R J S and Gretarsson, A M and Grote, H and Grunewald, S and Guido, C J and Guo, X and Gushwa, K and Gustafson, E K and Gustafson, R and Hacker, J and Hall, E D and Hammond, G and Hanke, M and Hanks, J and Hanna, C and Hannam, M D and Hanson, J and Hardwick, T and Harry, G M and Harry, I W and Hart, M and Hartman, M T and Haster, C-J and Haughian, K and Hee, S and Heintze, M and Heinzel, G and Hendry, M and Heng, I S and Heptonstall, A W and Heurs, M and Hewitson, M and Hild, S and Hoak, D and Hodge, K A and Hollitt, S E and Holt, K and Hopkins, P and Hosken, D J and Hough, J and Houston, E and Howell, E J and Hu, Y M and Huerta, E and Hughey, B and Husa, S and Huttner, S H and Huynh, M and Huynh-Dinh, T and Idrisy, A and Indik, N and Ingram, D R and Inta, R and Islas, G and Isler, J C and Isogai, T and Iyer, B R and Izumi, K and Jacobson, M and Jang, H and Jawahar, S and Ji, Y and Jiménez-Forteza, F and Johnson, W W and Jones, D I and Jones, R and Ju, L and Haris, K and Kalogera, V and Kandhasamy, S and Kang, G and Kanner, J B and Katsavounidis, E and Katzman, W and Kaufer, H and Kaufer, S and Kaur, T and Kawabe, K and Kawazoe, F and Keiser, G M and Keitel, D and Kelley, D B and Kells, W and Keppel, D G and Key, J S and Khalaidovski, A and Khalili, F Y and Khazanov, E A and Kim, C and Kim, K and Kim, N G and Kim, N and Kim, Y.-M and King, E J and King, P J and Kinzel, D L and Kissel, J S and Klimenko, S and Kline, J and Koehlenbeck, S and Kokeyama, K and Kondrashov, V and Korobko, M and Korth, W Z and Kozak, D B and Kringel, V and Krishnan, B and Krueger, C and Kuehn, G and Kumar, A and Kumar, P and Kuo, L and Landry, M and Lantz, B and Larson, S and Lasky, P D and Lazzarini, A and Lazzaro, C and Le, J and Leaci, P and Leavey, S and Lebigot, E O and Lee, C H and Lee, H K and Lee, H M and Leong, J R and Levin, Y and Levine, B and Lewis, J and Li, T G F and Libbrecht, K and Libson, A and Lin, A C and Littenberg, T B and Lockerbie, N A and Lockett, V and Logue, J and Lombardi, A L and Lormand, M and Lough, J and Lubinski, M J and Lück, H and Lundgren, A P and Lynch, R and Ma, Y and Macarthur, J and MacDonald, T and Machenschalk, B and MacInnis, M and Macleod, D M and Magaña-Sandoval, F and Magee, R and Mageswaran, M and Maglione, C and Mailand, K and Mandel, I and Mandic, V and Mangano, V and Mansell, G L and Márka, S and Márka, Z and Markosyan, A and Maros, E and Martin, I W and Martin, R M and Martynov, D and Marx, J N and Mason, K and Massinger, T J and Matichard, F and Matone, L and Mavalvala, N and Mazumder, N and Mazzolo, G and McCarthy, R and McClelland, D E and McCormick, S and McGuire, S C and McIntyre, G and McIver, J and McLin, K and McWilliams, S and Meadors, G D and Meinders, M and Melatos, A and Mendell, G and Mercer, R A and Meshkov, S and Messenger, C and Meyers, P M and Miao, H and Middleton, H and Mikhailov, E E and Miller, A and Miller, J and Millhouse, M and Ming, J and Mirshekari, S and Mishra, C and Mitra, S and Mitrofanov, V P and Mitselmakher, G and Mittleman, R and Moe, B and Mohanty, S D and Mohapatra, S R P and Moore, B and Moraru, D and Moreno, G and Morriss, S R and Mossavi, K and Mow-Lowry, C M and Mueller, C L and Mueller, G and Mukherjee, S and Mullavey, A and Munch, J and Murphy, D and Murray, P G and Mytidis, A and Nash, T and Nayak, R K and Necula, V and Nedkova, K and Newton, G and Nguyen, T and Nielsen, A B and Nissanke, S and Nitz, A H and Nolting, D and Normandin, M E N and Nuttall, L K and Ochsner, E and O’Dell, J and Oelker, E and Ogin, G H and Oh, J J and Oh, S H and Ohme, F and Oppermann, P and Oram, R and O’Reilly, B and Ortega, W and O’Shaughnessy, R and Osthelder, C and Ott, C D and Ottaway, D J and Ottens, R S and Overmier, H and Owen, B J and Padilla, C and Pai, A and Pai, S and Palashov, O and Pal-Singh, A and Pan, H and Pankow, C and Pannarale, F and Pant, B C and Papa, M A and Paris, H and Patrick, Z and Pedraza, M and Pekowsky, L and Pele, A and Penn, S and Perreca, A and Phelps, M and Pierro, V and Pinto, I M and Pitkin, M and Poeld, J and Post, A and Poteomkin, A and Powell, J and Prasad, J and Predoi, V and Premachandra, S and Prestegard, T and Price, L R and Principe, M and Privitera, S and Prix, R and Prokhorov, L and Puncken, O and Pürrer, M and Qin, J and Quetschke, V and Quintero, E and Quiroga, G and Quitzow-James, R and Raab, F J and Rabeling, D S and Radkins, H and Raffai, P and Raja, S and Rajalakshmi, G and Rakhmanov, M and Ramirez, K and Raymond, V and Reed, C M and Reid, S and Reitze, D H and Reula, O and Riles, K and Robertson, N A and Robie, R and Rollins, J G and Roma, V and Romano, J D and Romanov, G and Romie, J H and Rowan, S and Rüdiger, A and Ryan, K and Sachdev, S and Sadecki, T and Sadeghian, L and Saleem, M and Salemi, F and Sammut, L and Sandberg, V and Sanders, J R and Sannibale, V and Santiago-Prieto, I and Sathyaprakash, B S and Saulson, P R and Savage, R and Sawadsky, A and Scheuer, J and Schilling, R and Schmidt, P and Schnabel, R and Schofield, R M S and Schreiber, E and Schuette, D and Schutz, B F and Scott, J and Scott, S M and Sellers, D and Sengupta, A S and Sergeev, A and Serna, G and Sevigny, A and Shaddock, D A and Shahriar, M S and Shaltev, M and Shao, Z and Shapiro, B and Shawhan, P and Shoemaker, D H and Sidery, T L and Siemens, X and Sigg, D and Silva, A D and Simakov, D and Singer, A and Singer, L and Singh, R and Sintes, A M and Slagmolen, B J J and Smith, J R and Smith, M R and Smith, R J E and Smith-Lefebvre, N D and Son, E J and Sorazu, B and Souradeep, T and Staley, A and Stebbins, J and Steinke, M and Steinlechner, J and Steinlechner, S and Steinmeyer, D and Stephens, B C and Steplewski, S and Stevenson, S and Stone, R and Strain, K A and Strigin, S and Sturani, R and Stuver, A L and Summerscales, T Z and Sutton, P J and Szczepanczyk, M and Szeifert, G and Talukder, D and Tanner, D B and Tápai, M and Tarabrin, S P and Taracchini, A and Taylor, R and Tellez, G and Theeg, T and Thirugnanasambandam, M P and Thomas, M and Thomas, P and Thorne, K A and Thorne, K S and Thrane, E and Tiwari, V and Tomlinson, C and Torres, C V and Torrie, C I and Traylor, G and Tse, M and Tshilumba, D and Ugolini, D and Unnikrishnan, C S and Urban, A L and Usman, S A and Vahlbruch, H and Vajente, G and Valdes, G and Vallisneri, M and van Veggel, A A and Vass, S and Vaulin, R and Vecchio, A and Veitch, J and Veitch, P J and Venkateswara, K and Vincent-Finley, R and Vitale, S and Vo, T and Vorvick, C and Vousden, W D and Vyatchanin, S P and Wade, A R and Wade, L and Wade, M and Walker, M and Wallace, L and Walsh, S and Wang, H and Wang, M and Wang, X and Ward, R L and Warner, J and Was, M and Weaver, B and Weinert, M and Weinstein, A J and Weiss, R and Welborn, T and Wen, L and Wessels, P and Westphal, T and Wette, K and Whelan, J T and Whitcomb, S E and White, D J and Whiting, B F and Wilkinson, C and Williams, L and Williams, R and Williamson, A R and Willis, J L and Willke, B and Wimmer, M and Winkler, W and Wipf, C C and Wittel, H and Woan, G and Worden, J and Xie, S and Yablon, J and Yakushin, I and Yam, W and Yamamoto, H and Yancey, C C and Yang, Q and Zanolin, M and Zhang, Fan and Zhang, L and Zhang, M and Zhang, Y and Zhao, C and Zhou, M and Zhu, X J and Zucker, M E and Zuraw, S and Zweizig, J},
	month = mar,
	year = {2015},
	note = {Publisher: IOP Publishing},
	pages = {074001},
}

@phdthesis{Gray:2021thesis,
  author       = {Rachel Gray},
  title        = {Gravitational Wave Cosmology: Measuring the Hubble Constant with Dark Standard Sirens},
  school       = {University of Glasgow},
  year         = {2021},
  address      = {Glasgow, United Kingdom},
  type         = {PhD thesis},
  url          = {https://theses.gla.ac.uk/82438/}
}

@ARTICLE{chen_2_2018,
	title = {A 2 per cent {Hubble} constant measurement from standard sirens within 5 years},
	volume = {562},
	issn = {0028-0836, 1476-4687},
	url = {http://arxiv.org/abs/1712.06531},
	doi = {10.1038/s41586-018-0606-0},
	number = {7728},
	urldate = {2025-06-24},
	journal = {\nat},
	author = {Chen, Hsin-Yu and Fishbach, Maya and Holz, Daniel E.},
	month = oct,
	year = {2018},
	note = {arXiv:1712.06531 [astro-ph]},
	pages = {545--547},
}

@ARTICLE{barbary2016,
       author = {{Barbary}, Kyle},
        title = "{SEP: Source Extractor as a library}",
      journal = {The Journal of Open Source Software},
         year = 2016,
        month = oct,
       volume = {1},
       number = {6},
          eid = {58},
        pages = {58},
          doi = {10.21105/joss.00058},
       adsurl = {https://ui.adsabs.harvard.edu/abs/2016JOSS....1...58B}
}

\bibliographystyle{aasjournalv7}

\end{document}